\newcommand{\beas}{\begin{eqnarray*}}
\newcommand{\eeas}{\end{eqnarray*}}

\newcommand{\diracslash}[1]{#1\llap{/\kern2pt}}

\newcommand{\be}{\begin{equation}}
\newcommand{\ee}{\end{equation}}
\newcommand{\bea}{\begin{eqnarray}}
\newcommand{\eea}{\end{eqnarray}}
\newcommand{\ba}[1]{\begin{array}{#1}}
\newcommand{\ea}{\end{array}}

\documentclass[preprint,prd,aps,floats,nofootinbib,showpacs,floatfix]{revtex4}
\usepackage{graphicx}
\begin{document}
%

%

\title{$J/\psi$ and $\eta_{c}$ masses in isospin asymmetric hot nuclear matter\\-- a QCD sum rule approach}

\author{Arvind Kumar}
\email{iitd.arvind@gmail.com}
\affiliation{Department of Physics, Indian Institute of Technology, Delhi,
Hauz Khas, New Delhi -- 110 016, India}

\author{Amruta Mishra}
\email{amruta@physics.iitd.ac.in,mishra@th.physik.uni-frankfurt.de}
\affiliation{Department of Physics, Indian Institute of Technology, Delhi,
Hauz Khas, New Delhi -- 110 016, India}

\begin{abstract}
We study the in-medium masses of the charmonium states $J/\psi$ 
and $\eta_{c}$ in the nuclear medium using QCD sum rule approach.
These mass modifications arise due to modifications of the scalar 
and the twist-2 gluon condensates in the hot hadronic matter. 
The scalar gluon condensate,
$\left\langle \frac{\alpha_{s}}{\pi} G^a_{\mu\nu} {G^a}^{\mu\nu} 
\right\rangle$ and the twist-2 tensorial gluon operator, 
$\left\langle  \frac{\alpha_{s}}{\pi} G^a_{\mu\sigma}
{{G^a}_\nu}^{\sigma} \right\rangle $ in the nuclear medium 
are calculated from the medium modification of a scalar dilaton field 
introduced to incorporate trace anomaly of QCD within the chiral SU(3) 
model used in the present investigation. The effects of isospin asymmetry, 
density and temperature of the nuclear medium on the in-medium masses 
of the lowest charmonium states $J/\psi$ and $\eta_{c}$ mesons are 
investigated in the present work. The results of the present 
investigation are compared with the existing results on the 
masses of these states. The medium modifications of the masses 
of these charmonium states ($J/\psi$ and $\eta_c$) seem to be 
appreciable at high densities and should modify the experimental
observables arising from the compressed baryonic matter produced 
in asymmetric heavy ion collision experiments at the future facility 
of FAIR, GSI.
\end{abstract}

\pacs{24.10.Cn; 13.75.Jz; 25.75.-q}
\maketitle

\def\bfm#1{\mbox{\boldmath $#1$}}

\maketitle

\section{Introduction}
\label{intro}
The study of in-medium hadronic properties is of considerable interest, 
both experimentally and theoretically in the present day strong interaction 
physics. The study of the in-medium properties of hadrons has direct 
relevance in the experiments where hadronic matter is probed at high 
densities and/or temperatures. The CBM experiment at FAIR, GSI is the 
one where the dense matter at high densities and moderate temperatures 
is planned to be produced. The medium modifications of the strange and
charm mesons and their effects on the experimental observables
are amongst the topics which are intended to be studied
extensively in these experiments. Therefore, the topic of 
study of charm mesons in the medium has gotten considerable interest
in the recent past. The medium modifications of the properties 
of the charm mesons, $D$ and $\bar{D}$ 
as well as the excited charmonium states can have important consequences 
on the production of open charm and the suppression of the $J/\psi$ in 
the heavy-ion collision experiments. The suppression of $J/\psi$ in the 
heavy-ion collisions may lead to the signature of the quark-gluon-plasma 
(QGP) \cite{blaiz,satz}. Also it is observed that the effect of hadron 
absorption of $J/\psi$ is not negligible \cite{zhang,brat5,elena}. 
In Ref.\cite{vog}, it was reported that the charmonium suppression 
observed in Pb + Pb collisions of NA50 experiment cannot be simply explained
by nucleon absorption, but needs some additional density dependent suppression 
mechanism. It was suggested in these studies that the comover scattering
\cite{vog,capella,cassing} can explain the additional suppression of 
charmonium. An important difference between $J/\psi$ suppression pattern 
in comovers interaction model and in a deconfining scenario is that, in the 
former case, the anomalous suppression sets in smoothly from peripheral 
to central collisions rather than in a sudden way when the deconfining 
threshold is reached \cite{capella}.
The $J/\psi$ suppression in nuclear collisions at SPS energies has been 
studied in covariant transport approach HSD in Ref.\cite{cassing}. The 
calculations show that the absorption of $J/\psi$'s by both nucleons and 
produced mesons can explain reasonably not only the total $J/\psi$ 
cross--section but also the transverse energy dependence of $J/\psi$ 
suppression measured in both proton-nucleus and nucleus collisions. 
In Ref.\cite{wang}, the cross section of $J/\psi$ dissociation
by gluons is used to calculate the $J/\psi$ suppression in an equilibrating
parton gas produced in high energy nuclear collisions. The large average
momentum in the hot gluon gas enables gluons to break up the $J/\psi$, 
while hadron matter at reasonable temperature does not provide sufficiently 
hard gluons. The multigluon exchange can lead to an attractive potential 
between a $c\bar{c}$ meson and a nucleon, such that, for example, 
the $\eta_{c}$ could form bound states even with light nuclei 
\cite{klingl,brodsky}.

The $D$ $(\bar{D})$ mesons are made up of light (u or d) antiquark (quark) 
and one heavy charm quark (charm antiquark). In the QCD sum rule calculations, 
the mass modifications of $D (\bar{D})$ mesons in the nuclear medium 
arise due to interactions of light antiquark (quark) present in the 
$D(\bar{D})$ mesons with the light quark condensate \cite{haya1,friman}. 
There is appreciable change in the light quark condensate in the 
nuclear medium and hence $D(\bar{D})$ meson mass, due to its 
interaction with the light quark condensate, change appreciably 
in the hadronic matter. The medium modifications of the $D$ mesons 
modify the decay widths of the charmonium states, which have been studied
in Ref. \cite{friman}. The charmonium states are made up of a 
heavy charm quark and a charm antiquark. Within the QCD sum rule 
calculations, it is suggested that these heavy quarkonium states 
interact with the nuclear medium through
the gluon condensates \cite{klingl} unlike the interaction of the  
light vector mesons with the nuclear medium which is through the light
quark condensates \cite{hatsuda}. This is because all the heavy quark 
condensates can be related to the gluon condensates via heavy-quark 
expansion \cite{kimlee}. Also in the nuclear medium there are no 
valence charm quarks to leading order in density and any interaction 
with the medium is gluonic. The medium modifications of the gluon 
condensates are seen to be small and this leads to the mass modifications
of $J/\psi$ and $\eta_{c}$ mesons, which are the lowest charmonium states 
to be small in the nuclear medium \cite{klingl}. The leading order 
perturbative calculations \cite{pes1} of the study of the charmonium
states also shows that the mass of $J/\psi$ is reduced slightly in the
nuclear medium. In Ref. \cite{lee1}, the mass modifications of the charmonium
states have been studied using QCD second order Stark effect and the linear 
density approximation for the gluon condensate in the nuclear medium.
This shows a small drop for the $J/\psi$ mass at the nuclear matter 
density, but there is seen to be significant shift in the masses of 
the excited states of charmonium ($\psi(3686)$ and $\psi(3770)$).
Using QCD second order Stark effect, the masses of the charmonium 
states were also studied Ref. \cite{amarvind} in the asymmetric nuclear 
medium at finite temperatures. These medium modifications were
investigated by computing the scalar gluon condensate in the hot
nuclear medium from the medium modification of a scalar dilaton 
field within a chiral SU(3) model  which was introduced to incorporate 
broken scale invariance of QCD. This investigation showed small drop
of the $J/\psi$ mass in the medium, whereas the masses of the excited
charmonium states are observed to have appreciable drop at high densities.

In the present investigation, we study the in-medium modifications of 
the vector meson, $J/\psi$ and the pseudoscalar meson, $\eta_{c}$, 
using QCD sum rules \cite{klingl} and an effective chiral $SU(3)$ model 
\cite{papa}. 
To apply the QCD sum rules for the study of in-medium modifications 
of $J/\psi$ and $\eta_{c}$ mesons, we consider the contributions of 
the scalar gluon condensates, $\left\langle \frac{\alpha_{s}}{\pi}
G^a_{\mu\nu} {G^a}^{\mu\nu} \right\rangle$ and twist-2 tensorial gluon 
operator, $\left\langle  \frac{\alpha_{s}}{\pi} G^a_{\mu\sigma}
{{G^a}_{\nu}}^{\sigma} \right\rangle $ upto dimension four \cite{klingl}. 
The scalar gluon condensate as well as the twist-2 gluon operator
in the nuclear medium are calculated from the medium 
modification of a scalar dilaton field, $\chi$, introduced within 
a chiral $SU(3)$ model \cite{papa} through a scale symmetry breaking 
term in the Lagrangian density leading to the QCD trace anomaly. 
The chiral $SU(3)$ model \cite{papa} has been used successfully 
to study the medium modifications of kaons and antikaons in isospin 
asymmetric nuclear matter in \cite{isoamss} and in hyperonic matter 
in \cite{isoamss2}. The chiral $SU(3)$ model was generalized to $SU(4)$ 
to study the mass modifications of $D$-mesons arising from their 
interactions with the light hadrons in isospin symmetric hot hadronic 
matter in Ref.\cite{dmeson} and in isospin asymmetric nuclear matter 
at zero temperature \cite{amarind} and finite temperatures \cite{amarvind}
respectively. The in-medium properties of the vector mesons have also been 
studied within the model \cite{hartree,kristof1}. In the present 
investigation, we study the in-medium masses of the $J/\psi$ and 
$\eta_{c}$ mesons, calculated from the medium modifications of the 
dilaton field, $\chi$, in the nuclear asymmetric nuclear matter 
at finite temperatures within the chiral SU(3) model. 

The outline of the paper is as follows : In section II, we give a 
brief introduction of chiral $SU(3)$ model used to study the 
in-medium masses of charmonium states $J/\psi$ and $\eta_c$,
in the present investigation. The medium modifications of 
these charmonium states, $J/\psi$ and $\eta_{c}$ mesons arise 
from the medium modification of the scalar gluon condensate in the 
nuclear medium, simulated by a scalar dilaton field introduced 
in the hadronic model to incorporate broken scale invariance 
of QCD leading to QCD trace anomaly and also due to the medium
modification of the expectation value of the twist-2 gluon operator. 
Section III discusses briefly the QCD sum rule approach used to 
calculate the masses of the charmonium states $J/\psi$ and $\eta_c$. 
In section IV, we discuss the results of the present investigation. 
Section V summarizes the conclusions of the present work. 
 
\section{ The hadronic chiral $SU(3) \times SU(3)$ model}
\label{sec:2}
We use an effective chiral $SU(3)$ model for the present investigation 
\cite{papa}. The model is based on the nonlinear realization of chiral 
symmetry \cite{weinberg,coleman,bardeen} and broken scale invariance 
\cite{papa,hartree,kristof1}. This model has been used successfully to 
describe nuclear matter, finite nuclei, hypernuclei and neutron stars. 
The effective hadronic chiral Lagrangian density contains the following 
terms
\begin{equation}
{\cal L} = {\cal L}_{kin}+\sum_{W=X,Y,V,A,u} {\cal L}_{BW} + 
{\cal L}_{vec} + {\cal L}_{0} + {\cal L}_{SB}
\label{genlag}
\end{equation}
In Eq. (\ref{genlag}), ${\cal L}_{kin}$ is kinetic energy term, 
${\cal L}_{BW}$ is the baryon-meson interaction term in which the 
baryon-spin-0 meson interaction term generates the vacuum baryon masses. 
${\cal L}_{vec}$  describes the dynamical mass generation of the vector 
mesons via couplings to the scalar mesons and contain additionally 
quartic self-interactions of the vector fields. ${\cal L}_{0}$ contains 
the meson-meson interaction terms inducing the spontaneous breaking of 
chiral symmerty as well as a scale invariance breaking logarthimic 
potential. ${\cal L}_{SB}$ describes the explicit chiral symmetry breaking. 

To study the hadron properties at finite temperature and densities
in the present investigation, we use the mean  field approximation,
where all the meson fields are treated as classical fields. 
In this approximation, only the scalar and the vector fields 
contribute to the baryon-meson interaction, ${\cal L}_{BW}$
since for all the other mesons, the expectation values are zero.
The interactions of the scalar mesons and vector mesons with the
baryons are given as
\begin{equation}
{\cal  L}_{Bscal} +  {\cal L} _{Bvec} = - \sum_{i} \bar {\psi_i} 
\left[ m_{i}^{*} + g_{\omega i} \gamma_{0} \omega 
+ g_{\rho i} \gamma_{0} \rho + g_{\phi i} \gamma_{0} \phi \right] \psi_{i}. 
\end{equation}
The interaction of the vector mesons, of the scalar fields and 
the interaction corresponding to the explicitly symmetry breaking
in the mean field approximation are given as
\begin{eqnarray}
 {\cal L} _{vec} & = & \frac {1}{2} \left( m_{\omega}^{2} \omega^{2} 
+ m_{\rho}^{2} \rho^{2} + m_{\phi}^{2} \phi^{2} \right) 
\frac {\chi^{2}}{\chi_{0}^{2}}
+  g_4 (\omega ^4 +6\omega^2 \rho^2+\rho^4 + 2\phi^4),
\end{eqnarray}
\begin{eqnarray}
{\cal L} _{0} & = & -\frac{1}{2} k_{0}\chi^{2} \left( \sigma^{2} + \zeta^{2} 
+ \delta^{2} \right) + k_{1} \left( \sigma^{2} + \zeta^{2} + \delta^{2} 
\right)^{2} \nonumber\\
&+& k_{2} \left( \frac {\sigma^{4}}{2} + \frac {\delta^{4}}{2} + 3 \sigma^{2} 
\delta^{2} + \zeta^{4} \right) 
+ k_{3}\chi\left( \sigma^{2} - \delta^{2} \right)\zeta \nonumber\\
&-& k_{4} \chi^{4} 
 -  \frac {1}{4} \chi^{4} {\rm {ln}} 
\frac{\chi^{4}}{\chi_{0}^{4}}
+ \frac {d}{3} \chi^{4} {\rm {ln}} \Bigg (\bigg( \frac {\left( \sigma^{2} 
- \delta^{2}\right) \zeta }{\sigma_{0}^{2} \zeta_{0}} \bigg) 
\bigg (\frac {\chi}{\chi_0}\bigg)^3 \Bigg ),
\label{lagscal}
\end{eqnarray}
and 
\begin{eqnarray}
{\cal L} _{SB} = - \left( \frac {\chi}{\chi_{0}}\right)^{2} 
\left[ m_{\pi}^{2} 
f_{\pi} \sigma
+ \big( \sqrt {2} m_{k}^{2}f_{k} - \frac {1}{\sqrt {2}} 
m_{\pi}^{2} f_{\pi} \big) \zeta \right]. 
\label{lsb}
\end{eqnarray}
The effective mass of the baryon of species $i$ is given as
\begin{equation}
{m_i}^{*} = -(g_{\sigma i}\sigma + g_{\zeta i}\zeta + g_{\delta i}\delta)
\label{mbeff}
\end{equation}
The baryon-scalar meson interactions, as can be seen from equation
(\ref{mbeff}), generate the baryon masses through 
the coupling of  baryons to the non-strange $\sigma$, the strange $\zeta$ 
scalar mesons and also to scalar-isovector meson $\delta$. In analogy 
to the baryon-scalar meson couplings, there exist two independent 
baryon-vector meson interaction terms corresponding to the F-type 
(antisymmetric) and D-type (symmetric) couplings. Here antisymmetric 
coupling is used because the universality principle \cite{saku69} 
and vector meson dominance model suggest small symmetric couplings. 
Additionally,  we choose the parameters \cite{papa,isoamss} so as 
to decouple the strange vector field $\phi_{\mu}\sim\bar{s}\gamma_{\mu}s$ 
from the nucleon, corresponding to an ideal mixing between $\omega$ and 
$\phi$ mesons. A small deviation of the mixing angle from ideal mixing 
\cite{dumbrajs,rijken,hohler1} has not been taken into account in the 
present investigation.

The concept of broken scale invariance leading to the trace anomaly 
in (massless) QCD, $\theta_{\mu}^{\mu} = \frac{\beta_{QCD}}{2g} 
{G^a}_{\mu\nu} G^{\mu\nu a}$, where $G_{\mu\nu}^{a} $ is the 
gluon field strength tensor of QCD, is simulated in the effective 
Lagrangian at tree level \cite{sche1} through the introduction of 
the scale breaking terms 
\begin{eqnarray}
{\cal L}_{\rm {scalebreaking}} & = &  -\frac{1}{4} \chi^{4} {\rm {ln}}
\Bigg ( \frac{\chi^{4}} {\chi_{0}^{4}} \Bigg )
+ \frac{d}{3}{\chi ^4} 
{\rm {ln}} \Bigg ( \bigg (\frac{I_{3}}{{\rm {det}}\langle X 
\rangle _0} \bigg ) \bigg ( \frac {\chi}{\chi_0}\bigg)^3 \Bigg ),
\label{scalebreak}
\end{eqnarray}
where $I_3={\rm {det}}\langle X \rangle$, with $X$ as the multiplet
for the scalar mesons. These scale breaking terms,
in the mean field approximation, are given by the last two terms
of the Lagrangian density, ${\cal L}_0$  given by equation (\ref{lagscal}) 
\cite{heide1}. 
Within the chiral SU(3) model used in the present investigation,
the scalar gluon condensate 
$ \langle \frac{\alpha_s}{\pi}  {G^a}_{\mu\nu} G^{\mu\nu a} \rangle $,
as well as the twist-2 gluon operator,
$ \langle \frac{\alpha_s}{\pi}  {G^a}_{\mu\sigma} 
{G_\nu}^{\sigma a} \rangle $,
are simulated by the scalar dilaton field, $\chi$. These are
obtained from the energy momentum tensor 
\begin{eqnarray}
T_{\mu \nu}=(\partial _\mu \chi) 
\Bigg (\frac {\partial {{\cal L}_\chi}}
{\partial (\partial ^\nu \chi)}\Bigg )
- g_{\mu \nu} {\cal L}_\chi,
\label{energymom}
\end{eqnarray}
derived from the Lagrangian density for the dilaton field, given as
\begin{eqnarray}
{\cal L}_\chi & = & \frac {1}{2} (\partial _\mu \chi)(\partial ^\mu \chi)
- k_4 \chi^4 \nonumber \\ & - & \frac{1}{4} \chi^{4} {\rm {ln}} 
\Bigg ( \frac{\chi^{4}} {\chi_{0}^{4}} \Bigg )
+ \frac {d}{3} \chi^{4} {\rm {ln}} \Bigg (\bigg( \frac {\left( \sigma^{2} 
- \delta^{2}\right) \zeta }{\sigma_{0}^{2} \zeta_{0}} \bigg) 
\bigg (\frac {\chi}{\chi_0}\bigg)^3 \Bigg ),
\label{lagchi}
\end{eqnarray}
In massless QCD, the energy momentum tensor can be written as
\cite{leemorita2,moritalee2008} 
\begin{eqnarray}
T_{\mu \nu}=-ST({G^a}_{\mu\sigma} {{G^a}_\nu}^{ \sigma})
+ \frac {g_{\mu \nu}}{4} \frac{\beta_{QCD}}{2g} 
G_{\sigma\kappa}^{a} {G^a}^{\sigma\kappa} 
\label{energymomqcd}
\end{eqnarray}
where the first term is the symmetric traceless part and second term
is the trace part of the energy momentum tensor.
Writing 
\begin{eqnarray}
\langle \frac {\alpha_s}{\pi}{G^a}_{\mu\sigma} {{G^a}_\nu}^{\sigma} \rangle
=\Big (u_\mu u_\nu - \frac{g_{\mu \nu}}{4} \Big ) G_2,
\label{twist2g2}
\end{eqnarray}
where $u_\mu$ is the 4-velocity of the nuclear medium,
taken as $u_\mu =(1,0,0,0)$, we obtain the energy momentum tensor
in QCD as 
\begin{eqnarray}
T_{\mu \nu}=-\Big (\frac{\pi}{\alpha_s}\Big)\Big (u_\mu u_\nu - 
\frac{g_{\mu \nu}}{4} \Big ) G_2
+ \frac {g_{\mu \nu}}{4} \frac{\beta_{QCD}}{2g} 
{G^a}_{\sigma \kappa} {G^a}^{\sigma \kappa} 
\label{energymomqcdg2}
\end{eqnarray}
Equating the energy-momentum tensors given by
equations (\ref{energymom}) and (\ref{energymomqcdg2}) and
multiplying by $(u^\mu u^\nu -\frac {g^{\mu \nu}}{4})$,
we obtain the expression for $G_2$ as 
\begin{eqnarray}
G_2= - \frac {\alpha_s}{\pi} \Bigg ( \Big ( \partial_\alpha \chi \Big )
\Big (\frac {\partial {{\cal L}_\chi}} {\partial (\partial _\alpha \chi)}
\Big )
+\frac {4}{3} (\partial _i \chi)(\partial _i \chi) \Bigg ).
\label{g2}
\end{eqnarray}
We might note here that by multiplying the energy momentum tensor
of QCD given by equation (\ref{energymomqcdg2}) by $(u^\mu u^\nu 
-\frac {g^{\mu \nu}}{4})$, we project out the traceless part 
given by the first term of the energy momentum tensor,
described by the function, $G_2$. This is because $g_{\mu \nu}(u^\mu u^\nu 
-\frac {g^{\mu \nu}}{4})=0$, and hence, there is no contribution
from the trace part of the energy momentum tensor in QCD,
when we multiply the same by $(u^\mu u^\nu -\frac {g^{\mu \nu}}{4})$.
Similarly, by multiplying the energy momentum tensor given by
equation (\ref{energymomqcdg2}) by $g^{\mu \nu}$, the first part
gives zero and only the second term contributes to the trace of 
the energy momentum tensor. 
The effect of the logarithmic terms in the chiral SU(3) model, given by
equation (\ref{lagchi}), is to break the scale invariance. Multiplying 
equation (\ref{energymom}) by $g^{\mu \nu}$, we obtain the trace of 
the energy momentum tensor within the chiral SU(3) model as
\begin{equation}
T_{\mu}^{\mu} = (\partial _\mu \chi) \Bigg (\frac {\partial {{\cal L}_\chi}}
{\partial (\partial _\mu \chi)}\Bigg ) -4 {{\cal L}_\chi}. 
\label{tensor}
\end{equation}
Using the Euler-Lagrange's equation for the $\chi$ field,
the trace of the energy momentum tensor in the chiral SU(3) model
can be expressed as \cite{amarvind,heide1}
\begin{equation}
T_{\mu}^{\mu} = \chi \frac{\partial {{\cal L}_\chi}}{\partial \chi} 
- 4{{\cal L}_\chi} = -(1-d)\chi^{4}.
\label{tensor1}
\end{equation}
Multiplying equation (\ref{energymomqcdg2}) by $g^{\mu \nu}$, we obtain
the trace of the energy momentum tensor in QCD as
\begin{equation}
T_{\mu}^{\mu} = \langle \frac{\beta_{QCD}}{2g} 
{G^a}_{\sigma \kappa} {G^a}^{\sigma \kappa} \rangle 
\label{tensor1qcd}
\end{equation}
Using the Euler-Lagrange's equation for $\chi$ and dropping a total divergence 
term in equation (\ref{g2}), the expression for $G_2$ can be written as
\begin{eqnarray}
G_2 & = & \frac{\alpha_s}{\pi}
\Bigg [\chi \frac {\partial {\cal L}_\chi}{\partial \chi} -\frac {4}{3}
(\partial_i \chi)(\partial _i \chi)\Bigg ]\nonumber \\
&=&  \frac{\alpha_s}{\pi}
\Bigg [-(1-d+4 k_4)\chi^4-\chi ^4 {\rm {ln}}
\Big (\frac{\chi^4}{{\chi_0}^4}\Big )\nonumber \\ 
& + & \frac {4}{3} d\chi^{4} {\rm {ln}} \Bigg (\bigg( \frac {\left( \sigma^{2} 
- \delta^{2}\right) \zeta }{\sigma_{0}^{2} \zeta_{0}} \bigg) 
\bigg (\frac {\chi}{\chi_0}\bigg)^3 \Bigg )
-\frac{4}{3} (\partial_i \chi)(\partial_i \chi) \Bigg ],
\label{g2total}
\end{eqnarray}
The twist-2 gluon operator has only contribution in the nuclear medium and 
is zero in vacuum \cite{klingl}. Hence $(G_2)_{vac}=0$, which implies that
\begin{eqnarray}
-(1-d+4 k_4){\chi_0}^4
-\frac{4}{3} \langle (\partial_i \chi)(\partial_i \chi) \rangle _{vac}=0
\label{g2vac}
\end{eqnarray}
Assuming the glueball field, $\chi$ to be non-relativistic, and hence
assuming that $\langle (\partial_i \chi) (\partial _i \chi)\rangle_{medium}
\simeq \langle (\partial_i \chi) (\partial _i \chi)\rangle_{vac}$
and using equation (\ref{g2vac}), the expression for $G_2$ is obtained 
from the equation (\ref{g2total}) as
\begin{eqnarray}
G_2 &=&  \frac{\alpha_s}{\pi}
\Bigg [-(1-d+4 k_4)(\chi^4-{\chi_0}^4)-\chi ^4 {\rm {ln}}
\Big (\frac{\chi^4}{{\chi_0}^4}\Big )\nonumber \\ 
& + & \frac {4}{3} d\chi^{4} {\rm {ln}} \Bigg (\bigg( \frac {\left( \sigma^{2} 
- \delta^{2}\right) \zeta }{\sigma_{0}^{2} \zeta_{0}} \bigg) 
\bigg (\frac {\chi}{\chi_0}\bigg)^3 \Bigg ) \Bigg ].
\label{g2approx}
\end{eqnarray}
The scalar gluon condensate and the twist-2 gluon operator, described
in terms of the function, $G_2$, given by equations (\ref{tensor1})
and (\ref{g2approx}) are thus related to the $\chi$ field,
which is solved from the coupled equations of motion of the
scalar fields within the chiral SU(3) model.
These medium dependent gluon condensates are then related

The coupled equations of motion for the non-strange scalar field $\sigma$, 
the strange scalar field $ \zeta$, the scalar-isovector field $ \delta$ 
and the dilaton field $\chi$, are derived from the Lagrangian density,
and are given as
\begin{eqnarray}
&& k_{0}\chi^{2}\sigma-4k_{1}\left( \sigma^{2}+\zeta^{2}
+\delta^{2}\right)\sigma-2k_{2}\left( \sigma^{3}+3\sigma\delta^{2}\right)
-2k_{3}\chi\sigma\zeta \nonumber\\
&-&\frac{d}{3} \chi^{4} \bigg (\frac{2\sigma}{\sigma^{2}-\delta^{2}}\bigg )
+\left( \frac{\chi}{\chi_{0}}\right) ^{2}m_{\pi}^{2}f_{\pi}
-\sum g_{\sigma i}\rho_{i}^{s} = 0 
\label{sigma}
\end{eqnarray}
\begin{eqnarray}
&& k_{0}\chi^{2}\zeta-4k_{1}\left( \sigma^{2}+\zeta^{2}+\delta^{2}\right)
\zeta-4k_{2}\zeta^{3}-k_{3}\chi\left( \sigma^{2}-\delta^{2}\right)\nonumber\\
&-&\frac{d}{3}\frac{\chi^{4}}{\zeta}+\left(\frac{\chi}{\chi_{0}} \right) 
^{2}\left[ \sqrt{2}m_{k}^{2}f_{k}-\frac{1}{\sqrt{2}} m_{\pi}^{2}f_{\pi}\right]
 -\sum g_{\zeta i}\rho_{i}^{s} = 0 
\label{zeta}
\end{eqnarray}
\begin{eqnarray}
& & k_{0}\chi^{2}\delta-4k_{1}\left( \sigma^{2}+\zeta^{2}+\delta^{2}\right)
\delta-2k_{2}\left( \delta^{3}+3\sigma^{2}\delta\right) +k_{3}\chi\delta 
\zeta \nonumber\\
& + &  \frac{2}{3} d \chi^4 \left( \frac{\delta}{\sigma^{2}-\delta^{2}}\right)
-\sum g_{\delta i}\rho_{i}^{s} = 0
\label{delta}
\end{eqnarray}
 
\begin{eqnarray}
& & k_{0}\chi \left( \sigma^{2}+\zeta^{2}+\delta^{2}\right)-k_{3}
\left( \sigma^{2}-\delta^{2}\right)\zeta + \chi^{3}\left[1
+{\rm {ln}}\left( \frac{\chi^{4}}{\chi_{0}^{4}}\right)  \right]
+(4k_{4}-d)\chi^{3}
\nonumber\\
& - & \frac{4}{3} d \chi^{3} {\rm {ln}} \Bigg ( \bigg (\frac{\left( \sigma^{2}
-\delta^{2}\right) \zeta}{\sigma_{0}^{2}\zeta_{0}} \bigg ) 
\bigg (\frac{\chi}{\chi_0}\bigg)^3 \Bigg ) 
+\frac{2\chi}{\chi_{0}^{2}}\left[ m_{\pi}^{2}
f_{\pi}\sigma +\left(\sqrt{2}m_{k}^{2}f_{k}-\frac{1}{\sqrt{2}}
m_{\pi}^{2}f_{\pi} \right) \zeta\right]  = 0 
\label{chi}
\end{eqnarray}
In the above, ${\rho_i}^s$ are the scalar densities for the baryons, 
given as 
\begin{eqnarray}
\rho_{i}^{s} = \gamma_{i}\int\frac{d^{3}k}{(2\pi)^{3}} 
\frac{m_{i}^{*}}{E_{i}^{*}(k)} 
\Bigg ( \frac {1}{e^{({E_i}^* (k) -{\mu_i}^*)/T}+1}
+ \frac {1}{e^{({E_i}^* (k) +{\mu_i}^*)/T}+1} \Bigg )
\label{scaldens}
\end{eqnarray}
where, ${E_i}^*(k)=(k^2+{{m_i}^*}^2)^{1/2}$, and, ${\mu _i}^* 
=\mu_i -g_{\omega i}\omega -g_{\rho i}\rho -g_{\phi i}\phi$, are the single 
particle energy and the effective chemical potential
for the baryon of species $i$, and,
$\gamma_i$=2 is the spin degeneracy factor \cite{isoamss}.

The above coupled equations of motion are solved to obtain the density 
and temperature dependent values of the scalar fields ($\sigma$,
$\zeta$ and $\delta$) and the dilaton field, $\chi$, in the isospin
asymmetric hot nuclear medium. As has been already mentioned, the value 
of the $\chi$ is related to the scalar gluon condensate 
as well as the twist-2 gluon operator in the hot 
hadronic medium, and is used to compute the in-medium masses of charmonium 
states, in the present investigation. The isospin asymmetry in the medium
is introduced through the scalar-isovector field $\delta$ 
and therefore the dilaton field obtained after solving the above 
equations is also dependent on the isospin asymmetry parameter,
$\eta$ defined as $\eta= ({\rho_n -\rho_p})/({2 \rho_B})$, 
where $\rho_n$ and $\rho_p$ are the number densities of the neutron
and the proton and $\rho_B$ is the baryon density. In the present 
investigation, we study the effect of isospin asymmetry of the medium 
on the masses of the charmonium states $J/\psi$ and $\eta_{c}$.

The comparison of the trace of the energy momentum tensor arising
from the trace anomaly of QCD with that of the present chiral model
given by equations (\ref{tensor1}) and (\ref{tensor1qcd}),
gives the relation of the dilaton field to the scalar gluon condensate.
We have, in the limit of massless quarks \cite{cohen},
\begin{equation}
T_{\mu}^{\mu} = \langle \frac{\beta_{QCD}}{2g} 
G_{\mu\nu}^{a} G^{\mu\nu a} \rangle  \equiv  -(1 - d)\chi^{4} 
\label{tensor2}
\end{equation}
In the case of finite quark masses, equation (\ref{tensor2})
gets modified to 
\begin{equation}
T_{\mu}^{\mu} = \sum_i m_i \bar {q_i} q_i+ \langle \frac{\beta_{QCD}}{2g} 
G_{\mu\nu}^{a} G^{\mu\nu a} \rangle  \equiv  -(1 - d)\chi^{4}, 
\label{tensor2m}
\end{equation}
where the first term of the energy-momentum tensor, within the chiral 
SU(3) model is the negative of the explicit chiral symmetry breaking
term, ${\cal L}_{SB}$ given by equation (\ref{lsb}).

The parameter $d$ in equation (\ref{tensor2m}) originates from the 
second logarithmic term of equation (\ref{scalebreak}). To get an 
insight into the value of the parameter 
$d$, we recall that the QCD $\beta$ function at one loop level, for 
$N_{c}$ colors and $N_{f}$ flavors is given by
\begin{equation}
\beta_{\rm {QCD}} \left( g \right) = -\frac{11 N_{c} g^{3}}{48 \pi^{2}} 
\left( 1 - \frac{2 N_{f}}{11 N_{c}} \right)  +  O(g^{5})
\label{beta}
\end{equation}
In the above equation, the first term in the parentheses arises from 
the (antiscreening) self-interaction of the gluons and the second term, 
proportional to $N_{f}$, arises from the (screening) contribution of 
quark pairs. For massless quarks, the equations (\ref{tensor2}) and 
(\ref{beta}) suggest the 
value of $d$ to be 6/33 for three flavors and three colors, and 
for the case of three colors and two flavors, the value of $d$ 
turns out to be 4/33, to be consistent with the one loop estimate 
of QCD $\beta$ function. These values give the order of magnitude 
about which the parameter $d$ can be taken \cite{heide1}, since one 
cannot rely on the one-loop estimate for $\beta_{\rm {QCD}}(g)$. 
In the present investigation of the in-medium properties of the 
charmonium states due to the medium modification of the dilaton 
field within chiral $SU(3)$ model, we use the value of $d$=0.064
\cite{amarind}. This parameter, along with the other parameters
corresponding to the  scalar Lagrangian density, ${\cal L}_0$ 
given by (\ref{lagscal}), are fitted so as to ensure 
extrema in the vacuum for the $\sigma$, $\zeta$ and $\chi$ field 
equations, to  reproduce the vacuum masses of the $\eta$ and $\eta '$ 
mesons, the mass of the $\sigma$ meson around 500 MeV, and,
pressure, p($\rho_0$)=0,
with $\rho_0$ as the nuclear matter saturation density \cite{papa,amarind}.

The trace of the energy-momentum tensor in QCD, using the 
one loop beta function given by equation (\ref{beta}),
for $N_c$=3 and $N_f$=3, and accounting for the finite quark masses
\cite{cohen} is given as,
\begin{equation}
T_{\mu}^{\mu} = - \frac{9}{8} \frac{\alpha_{s}}{\pi} 
{G^a}_{\mu\nu} {G^a}^{\mu\nu}
+\left( \frac {\chi}{\chi_{0}}\right)^{2} 
\left( m_{\pi}^{2} 
f_{\pi} \sigma
+ \big( \sqrt {2} m_{k}^{2}f_{k} - \frac {1}{\sqrt {2}} 
m_{\pi}^{2} f_{\pi} \big) \zeta \right). 
\label{tensor4}
\end{equation} 
Using equations (\ref{tensor2}) and (\ref{tensor4}), we can write  
\begin{equation}
\left\langle  \frac{\alpha_{s}}{\pi} {G^a}_{\mu\nu} {G^a}^{\mu\nu} 
\right\rangle =  \frac{8}{9} \Bigg [(1 - d) \chi^{4}
+\left( \frac {\chi}{\chi_{0}}\right)^{2} 
\left( m_{\pi}^{2} f_{\pi} \sigma
+ \big( \sqrt {2} m_{k}^{2}f_{k} - \frac {1}{\sqrt {2}} 
m_{\pi}^{2} f_{\pi} \big) \zeta \right) \Bigg ]. 
\label{chiglu}
\end{equation}
We thus see from the equation (\ref{chiglu}) that the scalar 
gluon condensate $\left\langle \frac{\alpha_{s}}{\pi} G_{\mu\nu}^{a} 
G^{\mu\nu a}\right\rangle$ is related to the dilaton field $\chi$.
For massless quarks, since the second term in (\ref{chiglu}) 
arising from explicit symmetry breaking is absent, the scalar 
gluon condensate becomes proportional to the fourth power of the 
dilaton field, $\chi$, in the chiral SU(3) model. 
As mentioned earlier, the in-medium masses of charmonium states are 
modified due to the scalar gluon condensate and the twist-2
gluon operators, which are calculated from the modification of 
the $\chi$ field.

\section{QCD sum rule approach and in-medium masses of $J/\psi$ and $\eta_c$}
\label{sec:3}
In the present section, we shall use the medium modifications
of the gluon condensate, calculated from the dilaton field
in the chiral effective model, to compute the masses of the
charmonium states $J/\psi$ and $\eta_c$ in isospin asymmetric 
hot nuclear matter. Using QCD sum rules \cite{klingl} 
the in-medium masses of the lowest charmonium states can be written as
\begin{equation}
m^{2} \simeq \frac{M_{n-1}^{J} (\xi)}{M_{n}^{J} (\xi)} - 4 m_{c}^{2} \xi
\label{masscharm}
\end{equation}
where $M_{n}^{J}$ is the $n$th moment of the meson and $\xi$ is the 
normalization scale.  Using operator product expansion, the moment 
$M_{n}^{J}$ can be written as \cite{klingl}
\begin{equation}
M_{n}^{J} (\xi) = A_{n}^{J} (\xi) \left[  1 + a_{n}^{J} (\xi) \alpha_{s} 
+ b_{n}^{J} (\xi) \phi_{b} + c_{n}^{J} (\xi) \phi_{c} \right],
\label{moment}  
\end{equation}
where $A_n^J(\xi)$, $a_n^J(\xi)$, $b_n^J (\xi)$ and $c_n^J (\xi)$
are the Wilson coefficients. The common factor $A_{n}^{J}$ results 
from the bare loop diagram. The coefficients $a_{n}^{J}$ take into 
account perturbative radiative corrections, while the coefficients 
$b_{n}^{J}$ are associated with the scalar gluon condensate term
\begin{equation}
\phi_{b} = \frac{4 \pi^{2}}{9} \frac{\left\langle \frac{\alpha_{s}}{\pi} 
G^a_{\mu \nu} {G^a}^{\mu \nu} \right\rangle }{(4 m_{c}^{2})^{2}} 
\label{phib} 
\end{equation}
As already mentioned, the contribution of the scalar gluon condensate 
is taken through the dilaton field within the chiral $SU(3)$ model
used in the present investigation. 
Using equation (\ref{chiglu}), the above equation can be rewritten
in terms of the dilaton field $\chi$, as
\begin{equation}
\phi_{b} =  \frac{32 \pi^{2}}{81(4 m_{c}^{2})^{2}}  
\Bigg [ (1-d)\chi^4 
+\left( \frac {\chi}{\chi_{0}}\right)^{2} 
\left( m_{\pi}^{2} f_{\pi} \sigma
+ \big( \sqrt {2} m_{k}^{2}f_{k} - \frac {1}{\sqrt {2}} 
m_{\pi}^{2} f_{\pi} \big) \zeta \right) \Bigg ]. 
\label{phibglu}
\end{equation} 
The coefficients $A_{n}^{J}, a_{n}^{J},$ and $b_{n}^{J}$ are listed in 
Ref.\cite{rein}. The coefficients $c_n^J$ are associated with the 
value of $\phi_{c}$, which gives the contribution from twist-2 gluon 
operator and is given as
\begin{equation}
\phi_{c} =  \frac{4 \pi^{2}}{3(4 m_{c}^{2})^{2}}G_2,  
\label{phicglu}
\end{equation}
where $G_2$ is given by equation (\ref{g2approx}).
We shall calculate the in-medium masses of the charmonium states $J/\psi$ and
$\eta_c$ in the hot asymmetric nuclear matter and shall compare the results
with the contribution from the  twist-2 gluon operator as calculated in the 
linear density approximation. In the low density approximation, the term 
$\phi_c$ is given as \cite{klingl} 
\begin{equation}
\phi_{c} = - \frac{2 \pi^{2}}{3} \frac{\left\langle \frac{\alpha_{s}}{\pi} 
A_{G} \right\rangle }{(4 m_{c}^{2})^{2}} m_{N} \rho_{B}.
\label{phiclin}
\end{equation}  
In the above equation, $A_{G}$ represents twice the momentum fraction 
carried by gluons in the nucleon and is set equal to 0.9 \cite{klingl}. 
$m_{N}$ and $\rho_{B}$ are the nucleon mass and baryon density respectively. 
The Wilson coefficients, $c_{n}^{J}$ in the vector channel (for $J/\psi$) and 
the pseudoscalar channel (for $\eta_{c}$) can be found in Ref. \cite{klingl}. 
The parameters $m_{c}$ and $\alpha_{s}$ are the running charm quark mass 
and running coupling constant and are $\xi$ dependent \cite{rein}. 
These are given by
\begin{equation}
\frac{m_{c} (\xi)}{m_{c}} = 1 - \frac{\alpha_{s}}{\pi} \left\lbrace 
\frac{2 + \xi}{1 + \xi} {\rm {ln}}(2 + \xi) - 2 {\rm {ln}}2\right\rbrace 
\label{mc}
\end{equation}
where, $m_{c} \equiv m_{c} (p^{2} = -m_{c}^{2}) = 1.26$ GeV \cite{rein2}, and
\begin{eqnarray}
\alpha_{s}\left( Q_{0}^{2} + 4 m_{c}^{2} \right)  & = &  
\alpha_{s}\left( 4 m_{c}^{2} \right)/\left( 1 + \frac{25}{12\pi} 
\alpha_{s}\left( 4 m_{c}^{2}\right)
{\rm {ln}} \frac{Q_{0}^{2} + 4 m_{c}^{2}}{4 m_{c}^{2}} \right) 
\end{eqnarray}
with, $\alpha_{s}\left( 4 m_{c}^{2} \right) \simeq 0.3$ and 
$Q_{0}^{2} = 4 m_{c}^{2}\xi$ \cite{rein}.

In the next section, we present and discuss the results of our present 
work of the investigation of the in-medium masses of $J/\Psi$ and $\eta_c$ 
in isospin asymmetric hot nuclear matter.

\section{Results and Discussions}
\label{sec:4}
In this section, we first investigate the effects of density, 
isospin-asymmetry and temperature of the nuclear medium on the dilaton 
field $\chi$ in the chiral SU(3) model, from which we obtain the 
expectation value of the scalar gluon condensate in the medium. 
Using the QCD sum rule approach, the in-medium masses of charmonium 
states  $J/\psi$ and $\eta_{c}$ are calculated from the medium dependence
of the gluon condensates. The medium dependent dilaton field, $\chi$
is obtained by solving the equations of motion of the scalar fields,
$\sigma$, $\zeta$, $\delta$ and $\chi$ given by equations (\ref{sigma}) 
to (\ref{chi}).
The values of the parameters used in the present investigation, 
are : $k_{0} = 2.54, k_{1} = 1.35, k_{2} = -4.78, k_{3} = -2.77$, 
$k_{4} = -0.22$ and $d =  0.064$, which are the parameters
occurring in the scalar meson interactions defined in equation 
(\ref{lagscal}). 
The vacuum values of the scalar isoscalar fields, $\sigma$ and $\zeta$ 
and the dilaton field $\chi$ are $-93.3$ MeV, $-106.6$ MeV and 409.77 MeV
respectively. 
The values, $g_{\sigma N} = 10.6$ and $g_{\zeta N} = -0.47$ are 
determined by fitting to the vacuum baryon masses. The other parameters 
fitted to the asymmetric nuclear matter saturation properties 
in the mean-field approximation are: $g_{\omega N}$ = 13.3, 
$g_{\rho p}$ = 5.5, $g_{4}$ = 79.7, $g_{\delta p}$ = 2.5, 
$m_{\zeta}$ = 1024.5 MeV, $ m_{\sigma}$ = 466.5 MeV 
and $m_{\delta}$ = 899.5 MeV. The nuclear matter saturation 
density used in the present investigation is $0.15$ fm$^{-3}$.

\begin{figure}
\includegraphics[width=18cm,height=18cm]{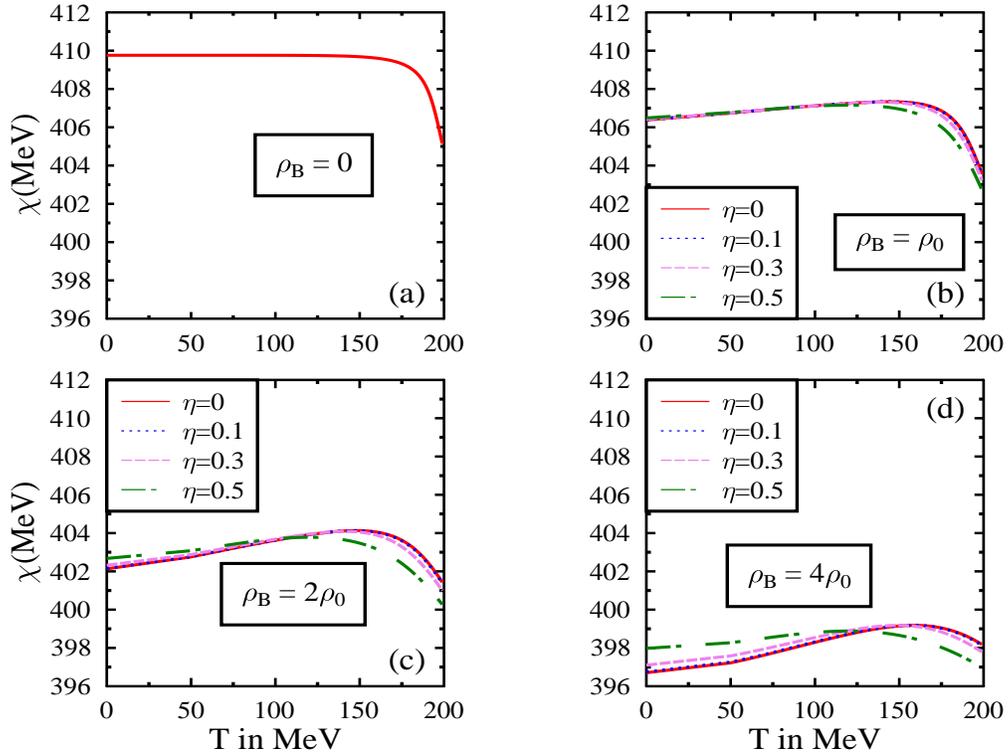}
\caption{(Color online) The dilaton field $\chi$ plotted as a function 
of the temperature, at given baryon densities, for different values of 
the isospin asymmetry parameter, $\eta$.}
\label{chitemp}
\end{figure}

In figure \ref{chitemp}, we show the variation of dilaton field $\chi$, 
with temperature, for both zero and finite baryon densities,
and for selected values of the isospin asymmetry parameter, 
$\eta$ = 0, 0.1, 0.3 and 0.5 \cite{amarvind}. At zero baryon density, 
it is observed 
that the value of the dilaton field remains almost 
a constant upto a temperature of about 130 MeV above which it is 
seen to drop with increase in temperature. However, the drop 
in the dilaton field is seen to be very small upto a temperature
of around 175 MeV above which the drop is seen to be larger. 
The value of the dilaton field is seen to change from 409.8 MeV 
at T=0 to about 409.7 MeV, 409.3 MeV and 405.76 MeV at T=150 MeV, 175 MeV
and 200 MeV respectively.
The thermal distribution functions have an effect of increasing 
the scalar densities at zero baryon density, i.e., for $\mu_i ^*$=0, 
as can be seen from the expression of the scalar densities, given 
by (\ref{scaldens}). This effect seems to be negligible upto
a temperature of about 130 MeV. This leads to a decrease in 
the magnitudes of scalar fields, $\sigma$ and $\zeta$. 
This behaviour of the scalar fields is reflected in the value of $\chi$, 
which is solved from the coupled equations of motion of the 
scalar fields, given by equations (\ref{sigma}), (\ref{zeta}), 
(\ref{delta}) and (\ref{chi}), as a drop as we 
increase the temperature above a temperature of about 130 MeV.
The scalar densities attaining nonzero values at high temperatures, 
even at zero baryon density, indicates the presence of baryon-antibaryon 
pairs in the thermal bath and has already been observed in the literature
\cite{kristof1,frunstl1}. This leads 
to the baryon masses to be different from their vacuum masses above this 
temperature, arising from modifications of the scalar fields $\sigma$ and
$\zeta$.

For finite density situations, the behaviour of the $\chi$ field
with temperature is seen to be very different from the zero density
case, as can be seen from the subplots (b),(c) and (d) of figure \ref{chitemp}, 
where the $\chi$ field is plotted as a function of the temperature
for densities $\rho_0$, 2$\rho_0$ and 4$\rho_0$ respectively. 
At finite densities, one observes first a rise and then a decrease of 
the dilaton field with temperature. This is related to the fact that 
at finite densities, the magnitude of the $\sigma$ field (as well as 
of the $\zeta$ field) first show an increase and then a drop with further 
increase of the temperature \cite{amarvind} which is reflected in the
behaviour of $\chi$ field, since it is solved from the coupled 
equations of the scalar fields. The reason for the different behaviour 
of the scalar fields ($\sigma$ and $\zeta$) at zero and finite densities
can be understood in the following manner \cite{kristof1}. As has
already been mentioned, the thermal distribution functions in (\ref{scaldens}) 
have an effect of increasing the scalar densities at zero baryon density, i.e., 
for $\mu_i ^*$=0. However, at finite densities, i.e., for nonzero values 
of the effective chemical potential, ${\mu_i}^*$, for increasing temperature, 
there are contributions also from higher momenta, thereby, 
increasing the denominator of the integrand on the right hand side of
the equation (\ref{scaldens}). This leads to a decrease in the scalar 
density. The competing effects of the thermal distribution functions
and the contributions of the higher momenta states
give rise to the observed effect of the scalar density
and hence of the $\sigma$ and $\zeta$ fields with temperature
at finite baryon densities \cite{kristof1}. This kind of behaviour 
of the scalar $\sigma$ field on temperature at finite densities 
has also been observed in the Walecka model by Li and Ko \cite{liko}, 
which was reflected as an increase in the mass of the nucleon with 
temperature at finite densities in the mean field calculations.
The effects of the behaviour of the scalar fields on the value 
of the $\chi$ field, obtained from solving the coupled equations 
(\ref{sigma}) to (\ref{chi}) for the scalar fields, are shown in 
figure \ref{chitemp}. 

In figure \ref{chitemp}, it is observed that for a given value of 
isospin asymmetry parameter $\eta$, the dilaton field $\chi$ decreases 
with increase in the density of the nuclear medium. The drop 
in the value of $\chi$ with density is seen to be much larger as compared
to its modification with temperature at a given density. 
For isospin symmetric nuclear medium ($\eta = 0$) at temperature $T = 0$, 
the reduction in the dilaton field $\chi$ from its vacuum value ($\chi_0$
 = 409.8 MeV), is seen to be about 3 MeV at $\rho_{B} = \rho_{0}$ 
and about 13 MeV, for $\rho_{B} = 4\rho_{0}$. 
As we move from isospin symmetric medium, with $\eta = 0$, to isospin 
asymmetric medium, at temperature $T = 0$, and, for a given value 
of density, there is seen to be an increase in the value of the 
dilaton field $\chi$. However, the effect of isospin asymmetry 
of the medium on the value of the dilaton field is observed to be
negligible upto about a density of nuclear matter saturation density, 
and is appreciable only at higher values of densities as can be seen 
in figure \ref{chitemp}. At nuclear matter saturation density, 
$\rho_{0}$, the value of dilaton field $\chi$ changes from 
$406.4$ MeV in symmetric nuclear medium ($\eta = 0$) to $406.5$ MeV in 
the isospin asymmetric nuclear medium ($\eta = 0.5$). At a density of 
about $4\rho_{0}$, the values of the dilaton field are modified to 
396.7 MeV and 398 MeV at $\eta =0$ and $0.5$, respectively. Thus the 
increase in the dilaton field $\chi$ with isospin asymmetry of the 
medium is seen to be more at zero temperature as we move to 
higher densities.

At a finite density, $\rho_{B}$, and for given isospin asymmetry 
parameter $\eta$, the dilaton field $\chi$ is seen to first increase 
with temperature and above a particular value of the temperature, 
it is seen to decrease with further increase in temperature. At the nuclear 
matter saturation density $\rho_{B} = \rho_{0}$ and in isospin symmetric 
nuclear medium ($\eta = 0$) the value of the dilaton field $\chi$ 
increases upto a temperature of about $T = 145$ MeV, above which there
is a drop in the dilaton field. For $\rho_B$=$\rho_0$ in the asymmetric 
nuclear matter with $\eta = 0.5$, there is seen to be a rise in the value 
of $\chi$ upto a temperature of about 120 MeV, above which it starts 
decreasing. As has already been mentioned, at zero temperature and for a 
given value of density, the dilaton field $\chi$ is found to increase 
with increase in the isospin asymmetry of the nuclear medium. But 
from figure \ref{chitemp}, it is observed that at high temperatures 
and for a given density, the value of the dilaton field $\chi$ becomes 
higher in symmetric nuclear medium as compared to isospin asymmetric 
nuclear medium e.g. at nuclear saturation density $\rho_{B} = \rho_{0}$ 
and temperature $T = 150$ MeV the values of dilaton field $\chi$ are 
$407.3$ MeV and $407$ MeV at $\eta = 0$ and $0.5$ respectively. 
At density $\rho_{B} = 4 \rho_{0}$, $T = 150$ MeV the values of 
dilaton field $\chi$ are seen to be $399.1$ MeV and $398.7$ MeV 
for $\eta = 0$ and $0.5$ respectively. This observed behaviour 
of the $\chi$ is related
to the fact that at finite densities and for isospin asymmetric matter, 
there are contributions from the scalar isovector $\delta$ field,
whose magnitude is seen to decrease for higher temperatures
for given densities, whereas $\delta$ field has zero contribution
for isospin symmetric matter.

\begin{table}
\begin{tabular}{||c||c||c||c||c||}
\hline
\multicolumn{1}{|c|} {}& \multicolumn{2}{|c|} {$J/\psi$} 
& \multicolumn{2}{|r|}  {$\eta_{c}\;\;\;\;\;\;\;\;\;\;\;$}  \\
\hline $\rho_{B}$ & $\eta$ = 0 & $\eta$ = 0.5 & $\eta$ = 0 & $\eta$ = 0.5 \\ 
\hline  $\rho_{0}$ & -4.48 & -4.34 & -5.21 & -5.06 \\ 
\hline  2$\rho_{0}$ & -10  & -9.29 & -9.14 & -8.64 \\ 
\hline  4$\rho_{0}$ & -16.77 & -15.19  & -13.12  & -12.19  \\ 
\hline 
\end{tabular}
\caption{The mass shifts of $J/\psi$ and $\eta_c$ are shown at densities
of $\rho_0$, 2$\rho_0$ and 4$\rho_0$ at values of the isospin asymmetric
parameter, $\eta$=0 and 0.5 for $\xi$=0.874. This value of $\xi$
reproduces the vacuum mass of $J/\psi$ as 3097 MeV.}
\label{table1} 
\end{table}

\begin{figure}
\includegraphics[width=18cm,height=18cm]{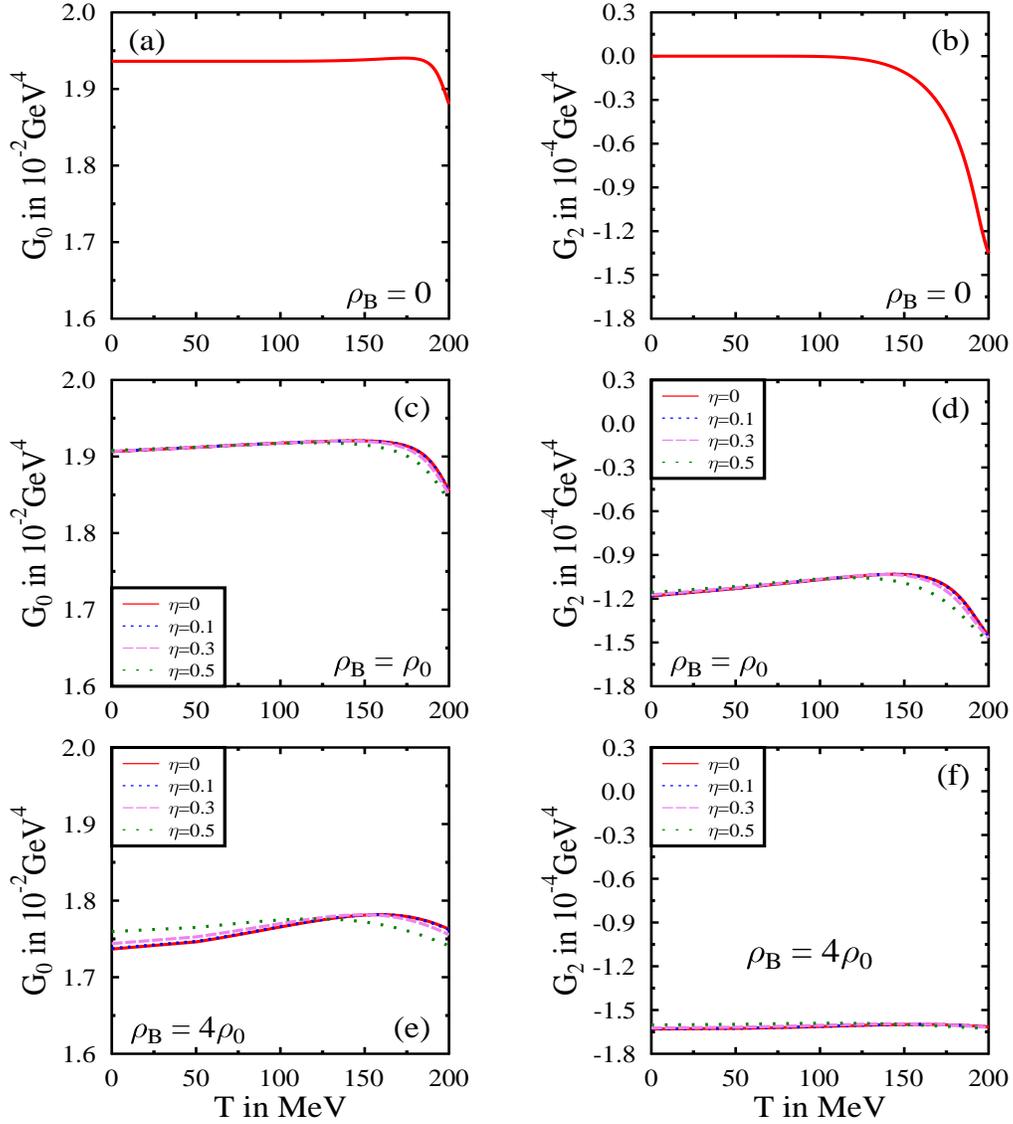} 
\caption{(Color online) The functions $G_0$ and $G_2$ describing
the trace and non-trace parts of the energy momentum tensor 
are plotted as functions of the density at different
temperatures and for different values of the isospin asymmetry 
parameter, $\eta$.}
\label{fig.2}

\end{figure}

In figure \ref{fig.2}, we show the variation of the trace and non-trace 
parts of the energy momentum tensor given by equation (\ref{energymomqcdg2}), 
with temperature, for different values of the baryon density and isospin 
asymmetry parameter, $\eta$. The trace part, $G_{0} = \left\langle 
\frac{\alpha_{s}}{\pi} G^{a}_{\mu\nu}G^{a\mu\nu} \right\rangle$, 
is given by equation (\ref{chiglu})
and $G_2$, which is related to the nontrace part of the 
energy momentum tensor is given by the equation (\ref{g2approx}),
both obtained from the SU(3) model used in the present investigation.
The value of the trace part, $G_{0}$ is plotted
as a function with temperature, for a densities, $\rho_B$=0, $\rho_0$ 
and 4$\rho_0$ in the subplots (a), (c) and (e) in figure \ref{fig.2}.
plotted in figure (\ref{fig.2}). For zero density, there is seen to be 
an increase of $G_0$ with temperature upto a temperature of about 175 MeV,
and then a drop with further increase in the temperature.
The values of $G_0$ are obtained as $1.9361 \times 10^{-2}$ GeV$^4$, 
$1.9362 \times 10^{-2}$ GeV$^4$, 
$1.9381 \times 10^{-2}$ GeV$^{4}$ and 1.88 $\times 10^{-2}$ GeV$^4$ 
at values of the temperature, T= 0, 100, 150 and 200 MeV respectively. 
We might note here that the calculations of the scalar gluon condensate, 
$G_0$, given by equation (\ref{chiglu}), have been performed by accounting 
for the finite quark 
masses in the trace anomaly. In the absence of finite quark masses,
the scalar gluon condensate becomes proportional to the fourth
power of the dilaton field, as can be seen from equation 
(\ref{chiglu}). The dilaton field is seen to decrease with temperature
at zero baryon density, as can be seen from figure \ref{chitemp}.
The values of $G_0$, for the limit of zero quark masses, also
decrease accordingly with temperature for $\rho_B$=0, with the
values of $G_0$ given as  $2.3455 \times 10^{-2}$ GeV$^4$, 
$2.34547 \times 10^{-2}$ GeV$^4$, $2.3437 \times 10^{-2}$ 
GeV$^{4}$ and 2.323 $\times 10^{-2}$ GeV$^4$  at values of 
temperature, T as 0, 100, 150 and 200 MeV respectively. 
A similar behaviour of $G_0$ with temperature
at zero baryon density has also been observed earlier in Ref.
\cite{moritalee2008}. In the present investigation,
the finite quark mass term leads to a decrease in the value of $G_0$, 
as can be seen from equation (\ref{chiglu}). At finite densities,
the dilaton field $\chi$ is seen to increase upto a temperature
above which it starts decreasing, as can be seen from figure 
\ref{chitemp}. Accounting for the finite quark masses,
we get a positive contribution to $G_0$ from the temperature
effects from the second term in equation (\ref{chiglu}), leading
to an increase in the scalar condensate upto a temperature 
above which there is seen to be a decrease with further rise 
in temperature.
For baryon densities of $\rho_B=\rho_0$ and $4\rho_0$, the values 
upto which $G_0$ increases with temperature are about 145 MeV and 
175 MeV respectively. In isospin symmetric nuclear matter,
for $\rho_B=\rho_0$, the values of $G_0$ are observed to be 
1.90646 $\times 10^{-2}$ GeV$^4$, 1.91755 $\times 10^{-2}$ GeV$^4$, 
1.92 $\times 10^{-2}$ GeV$^4$  and 1.8554 $\times 10^{-2}$ GeV$^4$ 
for temperatures of 0,100, 150 and 200 MeV respectively. For the same
values of the temperature, in the absence of finite quark masses,
the values of $G_0$ are observed to be 2.269 $\times 10^{-2}$ GeV$^4$, 
2.2857 $\times 10^{-2}$ GeV$^4$, 2.29 $\times 10^{-2}$ GeV$^4$
and 2.2 $\times 10^{-2}$ GeV$^4$ for $\rho_B=\rho_0$ and $\eta$=0.
In isospin symmetric nuclear matter for $\rho_B=4\rho_0$,
the values of $G_0$ are given as 1.7367 $\times 10^{-2}$ GeV$^4$ 
(2.06 $\times 10^{-2}$ GeV$^4$), 1.7656 $\times 10^{-2}$ GeV$^4$ 
(2.094 $\times 10^{-2}$ GeV$^4$), 1.78 $\times 10^{-2}$ GeV$^4$ 
(2.112 $\times 10^{-2}$ GeV$^4$) and 1.7626 $\times 10^{-2}$ GeV$^4$ 
(2.09 $\times 10^{-2}$ GeV$^4$) for values of temperature, T = 0, 100, 
150 and 200 MeV respectively, for the cases of the finite (zero) quark 
masses in the trace anomaly.

The non-trace part of the energy momentum tensor, $G_2$, is
plotted as a function of temperature in subplots (b), (d) and (f)
of figure \ref{fig.2} for densities, $\rho_B$=0, $\rho_0$ and 4$\rho_0$.
It may be noted that value of $G_{2}$ is zero in vacuum and this has 
a nonzero contribution only for finite density and/or temperature.
The magnitude of the quantity, $G_{2}$ is observed to increase with 
increase in the temperature of the nuclear medium for zero density,
with the values of $G_{2}$ at $\rho_{B} = 0$, given as
$ -7.106 \times 10^{-12}$ GeV$^4$, $-2.386 \times 10^{-7}$ GeV$^4$ and
$-1.106 \times 10^{-5}$ GeV$^{4}$ and $-3.528 \times 10^{-5}$ GeV$^4$
at values of temperature, T as 50, 100, 150  and 200 MeV respectively.
The observed behaviour of the magnitude of $G_{2}$ increasing  as a 
function of temperature at zero baryon density has also
been observed in Ref. \cite{moritalee2008}.
At nuclear saturation density, $\rho_{B} = \rho_{0}$
there is seen to be a decrease
in the magnitude of $G_2$ with temperature and then an increase 
with further rise in temperature.
In isospin symmetric medium, for $\rho_B=\rho_0$,  
the values of $G_{2}$ are given as $-1.181 \times 10^{-4}$ GeV$^4$,
 $-1.130 \times 10^{-4}$ GeV$^4$, $-1.06 9 \times 10^{-4}$ GeV$^4$, 
$-1.034 \times 10^{-4}$ GeV$^{4}$ and
$-1.4527 \times 10^{-4}$ GeV$^4$ at
values of temperature, T = 0, 50, 100, 150 and 200 MeV respectively.
For density $4\rho_{0}$ and $\eta$=0, the values of $G_{2}$ are given 
as $-1.63 \times 10^{-4}$ GeV$^4$, $-1.626 \times 10^{-4}$ GeV$^4$, 
$-1.613 \times 10^{-4}$ GeV$^4$, $-1.5992 \times 10^{-4}$ GeV$^{4}$
and $-1.6156 \times 10^{-4}$ GeV$^4$  for T = 0, 50, 100, 150
and 200 MeV respectively. In isospin asymmetric medium, $\eta = 0.5$,
at $\rho_{B} = 4\rho_{0}$, the values of $G_{2}$ are
$-1.602 \times 10^{-4}$ GeV$^4$, $-1.598 \times 10^{-4}$ GeV$^4$, 
$-1.591 \times 10^{-4}$ GeV$^4$,
$-1.5991 \times 10^{-4}$ GeV$^{4}$ and $-1.6265 \times 10^{-4}$ GeV$^4$
at temperature, T = 0, 50, 100, 150 and 200 MeV respectively.
In the present investigation, the effects of isospin asymmetry 
and temperature of the nuclear medium on the values of
$G_{0}$ and $G_{2}$ are observed to be small 
and the effect of density seems to be the dominant
effect. This is related to the fact that the dilaton field and
the scalar fields, $\sigma$, $\zeta$ and $\delta$ in the hot
isospin asymmetric nuclear medium are strongly dependent on the
density of the medium and the effects of temperature and isospin
asymmetry on these scalar fields are much smaller as compared
to the density effects. 

After obtaining the medium modification of the scalar gluon condensate
from the value of the dilaton field using equation (\ref{chiglu}),
and of the twist-2 gluon operator, by using equations (\ref{twist2g2})
and (\ref{g2approx}),
we next determine the in-medium mass shift of $J/\psi$ and $\eta_{c}$ mesons
using QCD sum rule approach. We use the moments in the range 
$5 \leq n \leq 12$
and fix the value of parameter $\xi = 0.874$, so that we can reproduce the 
vacuum value of mass of $J/\psi$, $m_{J/\psi}$ = $3097$ MeV. 
For this value of $\xi$, the parameters $\alpha_{s}$ and the running 
quark mass $m_{c}$ 
turn out to be $0.2667$ and $1.232 \times 10^3$ MeV respectively.
We consider the contributions 
from scalar gluon condensate $\left\langle \frac{\alpha_{s}}{\pi} 
G^a_{\sigma \kappa} {G^a}^{\sigma \kappa} \right\rangle$ and the
twist-2 tensorial gluon operator $\left\langle  \frac{\alpha_{s}}{\pi} 
G^a_{\mu \sigma} {{G^a}_\nu}^{\sigma}\right\rangle$ through the dilaton 
field $\chi$ within the chiral $SU(3)$ model used in the present
investigation. We obtain the value
of $\phi_b$ which is related to the scalar gluon condenstate by
equation (\ref{phib}), from the medium dependent $\chi$ field 
using the equation (\ref{phibglu}). The value for $\phi_c$ arising 
from the twist-2 tensorial gluon operator is calculated within the
chiral SU(3) model by using equation (\ref{phicglu}). 
We also compare our results with the twist-2 gluon operator as calculated
from the formula obtained in the low density 
approximation as given by equation (\ref{phiclin}) \cite{klingl}.
The value of $\phi_{c}$ at nuclear saturation density, $\rho_{0} = 
0.15$ fm$^{-3}$ is calculated to be $-4.2158 \times 10^{-5}$
within the SU(3) chiral model used in the present investigation,
which may be compared to the value of $-1.4685 \times 10^{-5}$
in the linear density approximation \cite{klingl}. 
In table \ref{table1}, we summarize the results for the mass shifts
of $J/\psi$ and $\eta_c$, as obtained in the present investigation, 
at zero temperature, for values of the baryon densities as $\rho_0$, 
$2\rho_0$ and $4\rho_0$, and for isospin asymmetry parameter as 
$\eta=0$ and 0.5, for the value of $\xi$=0.874. As has already been
mentioned this value of $\xi$ is fixed so as to obtain the 
observed vaccum mass of $J/\psi$ as 3097 MeV. 
\begin{figure}
\includegraphics[width=18cm,height=18cm]{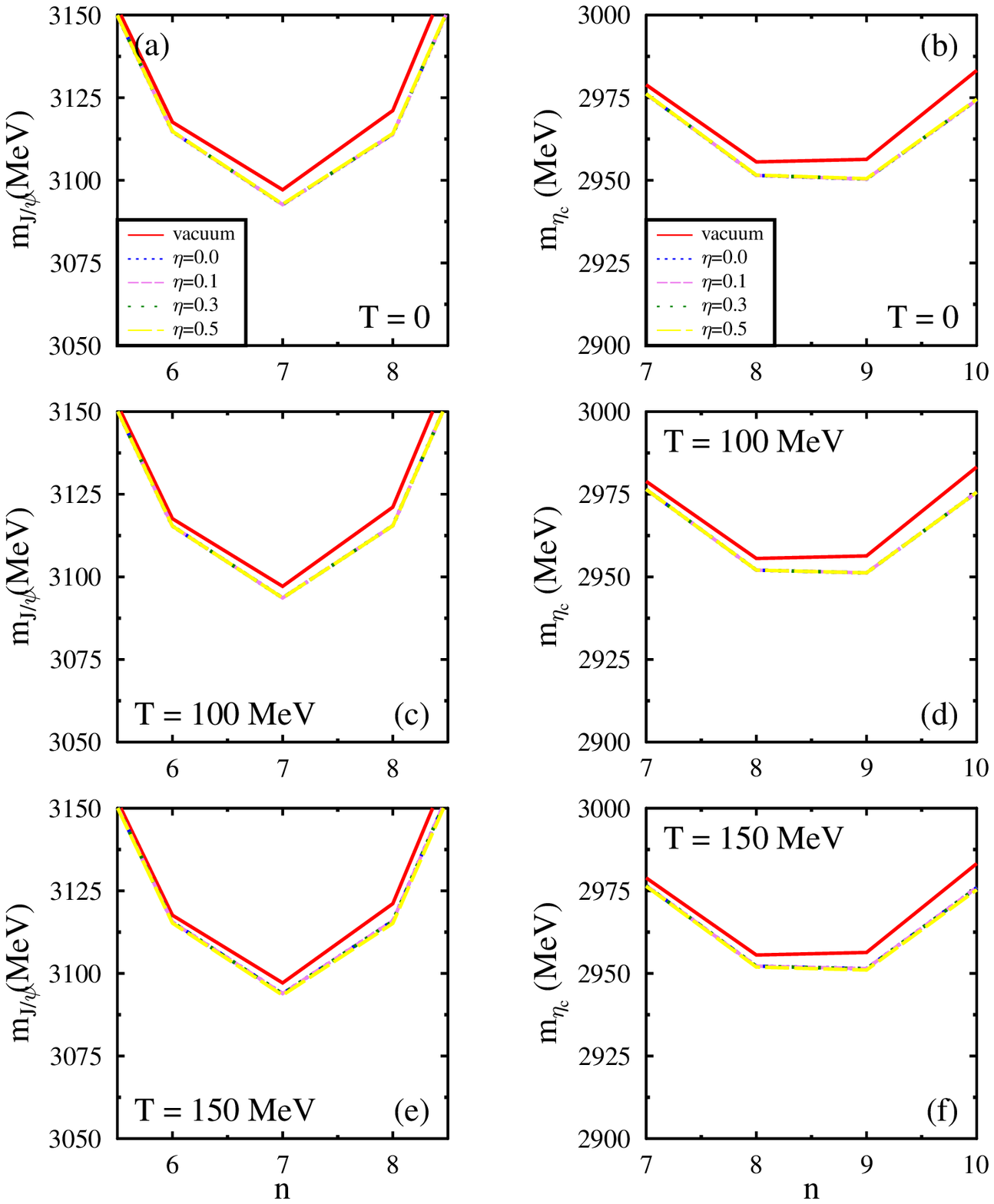} 
\caption{(Color online) The in-medium masses of the $J/\psi$ 
and $\eta_{c}$ mesons plotted as functions of $n$, for  
nuclear matter saturation density, $\rho_{0}$ at different
temperatures and for different values of the isospin asymmetry 
parameter, $\eta$. The value of parameter $\xi$ is taken as 
0.874 which reproduces the vacuum mass of $J/\psi$ as 3097 MeV.} 
\label{fig.3}
\end{figure} 

\begin{figure}
\includegraphics[width=18cm,height=18cm]{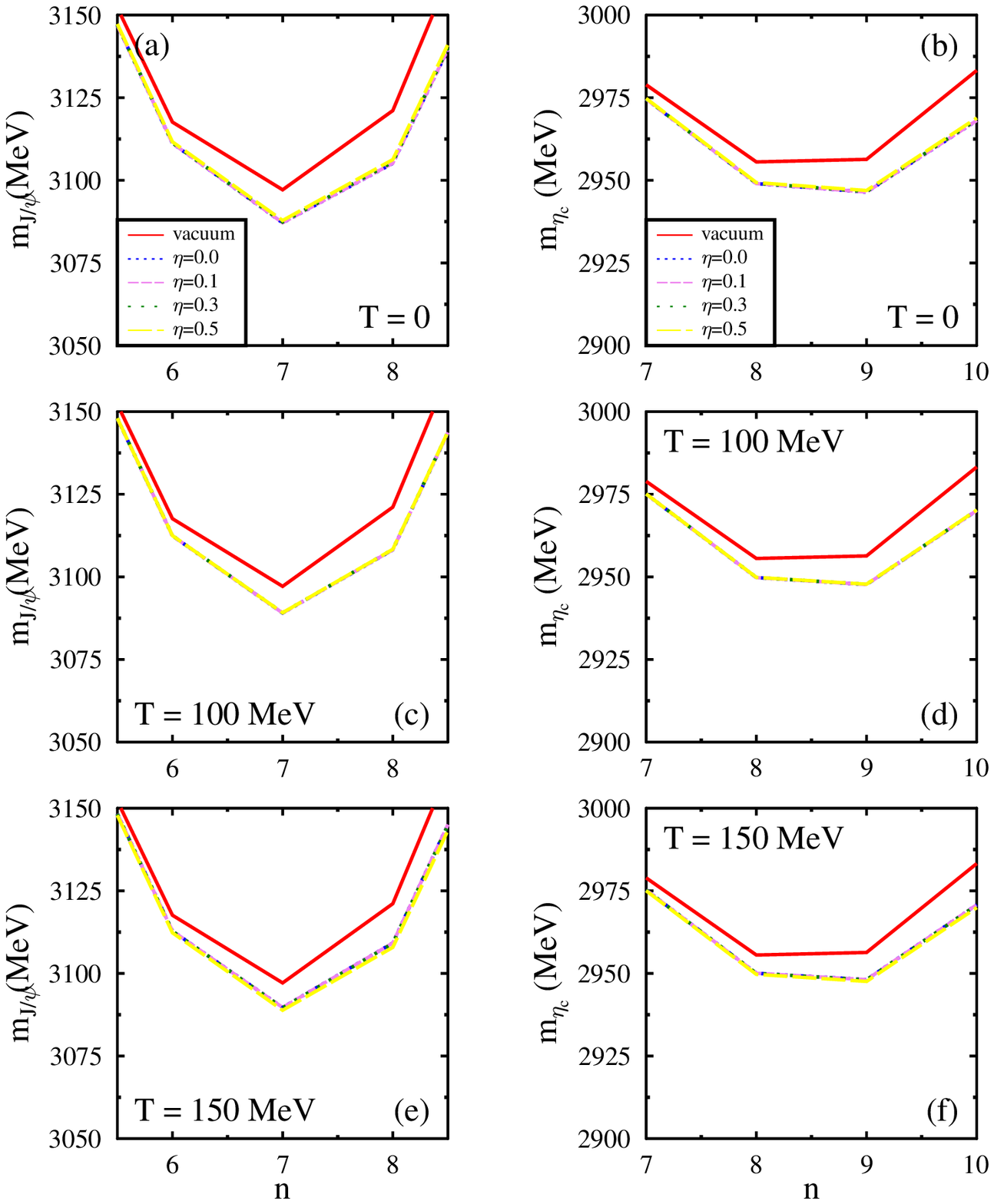} 
\caption{(Color online) The in-medium masses of the $J/\psi$ 
and $\eta_{c}$ mesons plotted as functions of $n$, for  
baryon density, $\rho_B=2\rho_{0}$ at different
temperatures and for different values of the isospin asymmetry 
parameter, $\eta$. The value of parameter $\xi$ is taken as 
0.874 which reproduces the vacuum mass of $J/\psi$ as 3097 MeV.} 
\label{fig.4}
\end{figure} 

In figure \ref{fig.3}, we show the variation of masses of $J/\psi$ and 
$\eta_{c}$ mesons with, n, for fixed value of baryon density, $\rho_{B}
 = \rho_{0}$, and for different values of isospin asymmetry parameter,
$\eta$. We show the results for values of temperature, $T = 0, 100$ 
and $150$ MeV. 
In symmetric nuclear matter, at nuclear matter saturation density, 
$\rho_{B} = \rho_{0}$, and at temperature $T = 0$, we obtain 
the mass shifts for $J/\psi$ and $\eta_{c}$ mesons to be equal to 
$-4.48$ MeV and $-5.21$ MeV respectively, as can be seen from
table \ref{table1}. These values of mass shifts for for $J/\psi$ 
and $\eta_{c}$ mesons may be compared with the mass shifts of 
$-7$ MeV and $-5$ MeV respectively obtained in the linear
density approximation in Ref.\cite{klingl}. In the present investigation, 
we calculate the values of $\phi_{b}$ and $\phi_c$ from the medium 
modification of the dilaton field, $\chi$, within the chiral $SU(3)$ 
model, by using equations (\ref{phibglu}) and (\ref{phicglu}).
In isospin symmetric nuclear medium, at baryon densities, $\rho_{B} = 0$ 
and $\rho_{0}$, the values of the dilaton field, $\chi$ are
$409.76$ and $406.38$ MeV respectively and hence using the equation 
(\ref{chiglu}), the values of the scalar gluon condensate 
$\left\langle \frac{\alpha_{s}}{\pi} G_{\mu\nu}^{a} {G^a}^{\mu\nu} 
\right\rangle$ turn out to be $(373$ MeV$)^{4}$ and 
$(371.6$ MeV$)^{4}$ for densities $\rho_B=0$ and $\rho_0$ 
respectively. We might note here that when we negelct the 
quark masses in the trace anomaly, the values of the 
scalar gluon condensate at these densities are modified to 
$(391\; {\rm {MeV}})^4$ and $(388 \;{\rm {MeV}})^4$ respectively. 
We thus observe an increase of the values of the scalar gluon
condensate by about 20\% when we do not account for the finite 
masses of the quarks. The values of $\phi_{b}$,
accounting for the finite quark masses,
turn out to be $2.3\times 10^{-3}$ and $2.27 \times 10^{-3}$ 
in the vacuum and at nuclear matter saturation 
density, $\rho_{0}$, respectively. These may be compared with
the values of $\phi_{b}$ to be equal to $1.7 \times 10^{-3}$ and 
$1.6 \times 10^{-3}$ respectively for $\rho_B=0$ and for 
$\rho_B=\rho_0$, obtained from the values of scalar gluon condensate 
of $(350 \;{\rm {MeV}})^{4}$ and $(344.81 \;{\rm {MeV}})^{4}$ respectively 
in vacuum and at nuclear saturation density, $\rho_{0}$
in Ref. \cite{klingl} in the linear density approximation. 
We might note here that the value of nuclear matter saturation 
density used in the present calculations is $0.15$ fm$^{-3}$ and 
in Ref. \cite{klingl}, it was taken to be $0.17$ fm$^{-3}$. 
When the quark masses are neglected, and $\phi_c$ as calculated
in the chiral SU(3) model used in the present investigation, 
the  values of the mass shift for $J/\psi$ and $\eta_c$ turn out 
to be -8.01 MeV and -5.13 MeV respectively.
When we calculate $\phi_b$ within the chiral SU(3) model,
but calculate the contribution of the twist-2 operator through
$\phi_c$ calculated in the linear density approximation 
given by equation (\ref{phiclin}) \cite{klingl}, we obtain
the mass shifts for $J/\psi$ and $\eta_c$ at $\rho_B=\rho_0$
for symmetric nuclear matter at zero temperature to be 
given as -2.88 MeV and -2.02 MeV respectively.
The value of the mass shift of $J/\psi$ of about -4.48 MeV at
the nuclear matter density $\rho_0$ in symmetric nuclear matter
at zero temperature obtained in the present investigation. 
may be compared to the values of 
the mass shift of -8 MeV obtained using QCD second order Stark effect
using the value of the scalar gluon condensate obtained using 
a linear density approximation \cite{lee1} as well as a value of
-8.6 MeV, when the scalar gluon condensate was obtained from
the expectation value of the scalar dilaton field in a chiral
SU(3) model \cite{amarvind}.
We observe in figure \ref{fig.3} that the isospin dependence
of the mass shifts of $J/\psi$ and $\eta_c$ are very small.
This is due to the fact that the dependence of $\chi$ on the isospin
asymmetry is very small, as can be seen from figure \ref{chitemp}.

\begin{figure}
\includegraphics[width=18cm,height=18cm]{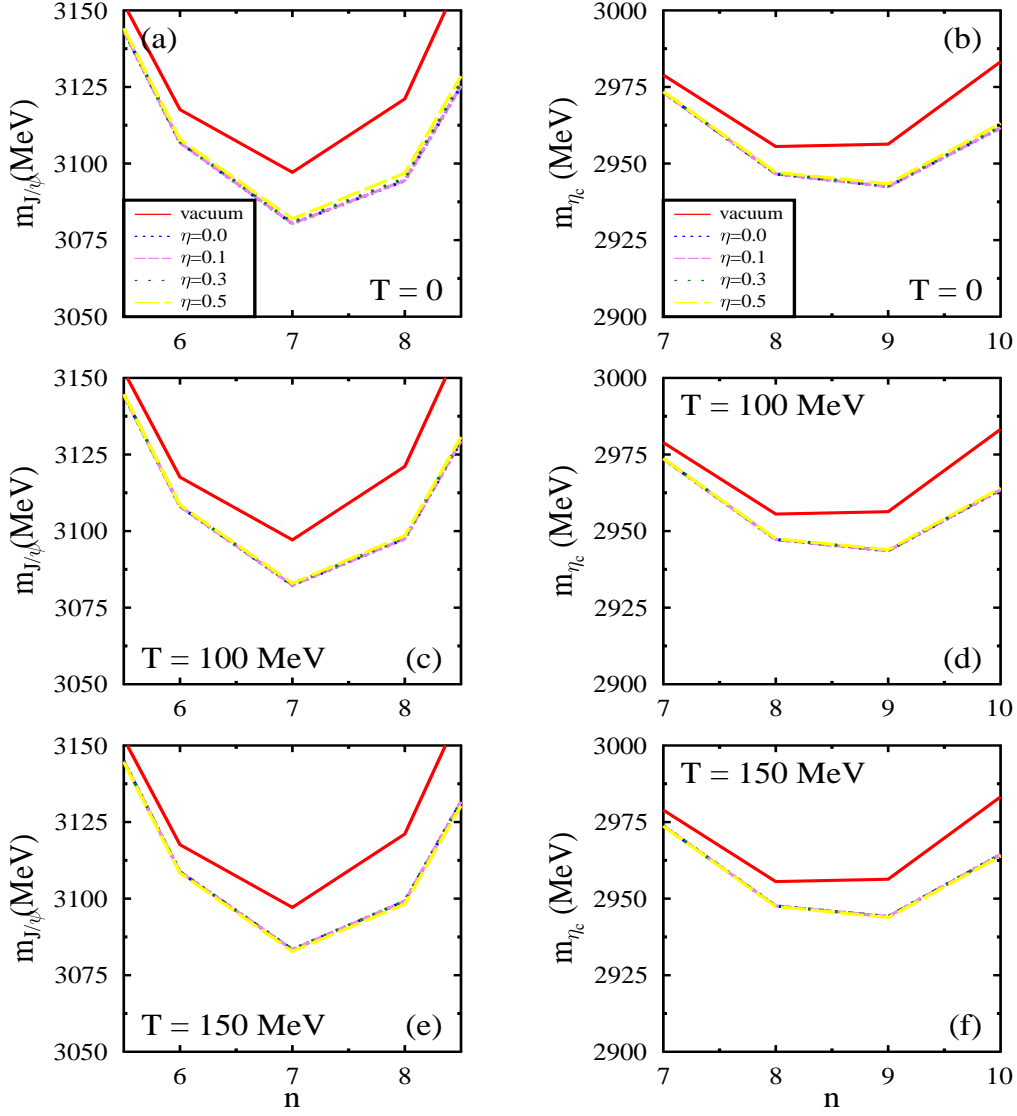} 
\caption{(Color online) The in-medium masses of the $J/\psi$ 
and $\eta_{c}$ mesons plotted as functions of $n$, for  
baryon density, $\rho_B=4\rho_{0}$ at different
temperatures and for different values of the isospin asymmetry 
parameter, $\eta$. The value of parameter $\xi$ is taken as 
0.874 which reproduces the vacuum mass of $J/\psi$ as 3097 MeV.} 
\label{fig.5}
\end{figure}

Figures \ref{fig.4} and \ref{fig.5} show the mass shifts of $J/\psi$ 
and $\eta_c$ for baryon densities $\rho_{B} = 2\rho_{0}$ and $4\rho_{0}$ 
respectively, at different temperatures and different values
of the isospin asymmetry parameter, $\eta$. 
In isospin symmetric nuclear medium, at density $\rho_{B} = 2 \rho_{0}$, 
temperature $T = 0$, the mass shifts for $J/\psi$ and $\eta_{c}$ mesons 
are observed to be $-10$ MeV and $-9.14$ MeV respectively. 
The effects of isospin 
asymmetry of the medium on the mass shift of the $J/\psi$ and $\eta_{c}$ 
mesons are seen to be almost negligible, as can be seen from table
\ref{table1}. This is due to the very small changes in the dilaton field 
with the isospin asymmetry of the medium, as can be seen from figure 
\ref{chitemp}. In isospin asymmetric nuclear 
medium ($\eta = 0.5$), at nuclear saturation density $\rho_{0}$, 
the mass shifts in $J/\psi$ and $\eta_{c}$ mesons at zero temperature 
are observed to be $-4.34$ MeV and $-5.06$ MeV from their vacuum values,
which may be compared with the values of $-4.48$ MeV and $-5.21$ MeV
respectively for the isospin symmetric nuclear matter. 
At values of the baryon density $\rho_{B}$ as 2$\rho_0$ and 4$\rho_{0}$, 
as can be seen from table \ref{table1}, the isospin dependence
of the mass shifts for $J/\psi$ and $\eta_{c}$ mesons are 
seen to be negligibly small.

The effects of temperature on the dilaton field $\chi$ is very small and 
this is reflected in the small change in the mass shifts of  $J/\psi$ and 
$\eta_{c}$ mesons with temperature $T$. In isospin symmetric nuclear medium 
($\eta = 0$), at nuclear saturation density $\rho_{B} = 
\rho_{0}$, the mass shifts for $J/\psi$ mesons, from their vacuum values, 
are observed to be $-4.01$, $-3.5$ and $-3.23$ MeV at temperatures $T = 50, 
100$ and $150$ MeV respectively. At baryon density, $\rho_{B} = 4\rho_{0}$, 
the values of mass shift for $J/\psi$ meson changes to $-16.13$, $-14.82$ 
and $-13.76$ MeV at temperatures $T = 50, 100$ and $150$ MeV respectively. 
The value of the mass shift obtained at finite value of temperature 
is observed to be smaller as compared to the zero temperature case. This is 
because, at finite value of baryon density $\rho_{B}$, the dilaton 
field $\chi$ increases with increase in the temperature of the nuclear 
medium, but the increase is very small. 
In isospin asymmetric nuclear medium, $\eta = 0.5$, at density $\rho_{B} 
= 4\rho_{0}$, the mass shifts for $J/\psi$ mesons, from their 
vacuum values are $-14.82$, $-14.16$ and $-14.36$ MeV at 
temperatures $T = 50, 100$ and $150$ MeV respectively. 

For the pseudoscalar meson $\eta_{c}$, the mass shifts at nuclear saturation 
density $\rho_{0}$, in nuclear medium with $\eta$=0 (0.5), are $-4.81 (-4.73)$, 
$-4.352 (-4.345)$ and $-4.1 (-4.54) $ MeV at temperatures $T = 50, 100$ 
and $150$ MeV respectively.
At density $\rho_{B} = 4\rho_{0}$, with $\eta$=0 (0.5), these values 
are modified to $-12.77 (-11.98)$, $-12.02 (-11.6)$ and $-11.41 (-11.73)$ 
MeV respectively for T=50, 100 and 150 MeV. 
It may be noted that at high values of temperatures e.g. 
at $T = 150$ MeV, the mass shift is more in the isospin asymmetric 
nuclear medium ($\eta = 0.5$) as compared to isospin symmetric nuclear 
medium ($\eta = 0$). This is opposite to the zero temperature case. 
The reason is that at high temperatures the dilaton field $\chi$ 
has larger drop in the isospin asymmetric nuclear medium ($\eta = 0.5$) 
as compared to the isospin symmetric nuclear medium ($\eta = 0$),
as can be seen in figure \ref{chitemp}.

\begin{table}
\begin{tabular}{||c||c||c||c||c||}
\hline
\multicolumn{1}{|c|} {}& \multicolumn{2}{|c|} {$J/\psi$} 
& \multicolumn{2}{|r|}  {$\eta_{c}\;\;\;\;\;\;\;\;\;\;\;$}  \\
\hline $\rho_{B}$ & $\eta$ = 0 & $\eta$ = 0.5 & $\eta$ = 0 & $\eta$ = 0.5 \\ 
\hline  $\rho_{0}$ & -4.27 & -4.14 & -5.69 & -5.54 \\ 
\hline  2$\rho_{0}$ & -9.55  & -8.87 & -9.39 & -8.92 \\ 
\hline  4$\rho_{0}$ & -16.02 & -14.51  & -13.12  & -12.25  \\ 
\hline 
\end{tabular} 
\caption{The mass shifts of $J/\psi$ and $\eta_c$ are shown at densities
of $\rho_0$, 2$\rho_0$ and 4$\rho_0$ at values of the isospin asymmetric
parameter, $\eta$=0 and 0.5 for $\xi$=0.8995. This value of $\xi$
reproduces the vacuum mass of $\eta_c$ as 2980.5 MeV.}
\label{table2}
\end{table}

As mentioned earlier, for the above calculations we had fixed the value of 
parameter $\xi$ so as to reproduce the vacuum value of $J/\psi$ mass. 
However, with this value of $\xi$, the vacuum value of $\eta_{c}$ meson 
comes out to be $2955.6$ MeV. 
We can reproduce the vacuum value of pseudoscalar meson $\eta_{c} 
= 2980.5$ MeV, if we fix the value of $\xi = 0.8995$. 
For this value of $\xi$, the parameters $\alpha_{s}$
and the running charm quark mass $m_{c}$ turn out to be 
$0.266$ and $1.2313 \times 10^3$ MeV respectively. 
For these values of parameters, the mass shifts for 
$J/\psi$ and $\eta_{c}$ mesons, in nuclear medium at zero temperature,
at densities $\rho_0$, 2$\rho_0$ and 4$\rho_0$ for $\eta$=0 (0.5)
are summarized in table \ref{table2}.

\begin{table}
\begin{tabular}{||l||l||l||l||r||}
\hline
\multicolumn{1}{|c|} {}& \multicolumn{2}{|c|} {$J/\psi$} 
& \multicolumn{2}{|r|}  {$\eta_{c}\;\;\;\;\;\;\;\;\;\;\;$}  \\
\hline $\rho_{B}$ & $\eta$ = 0 & $\eta$ = 0.5 & $\eta$ = 0 & $\eta$ = 0.5 \\ 
\hline  $\rho_{0}$ & -4.43 & -4.28 & -3.8 & -3.66 \\ 
\hline  2$\rho_{0}$ & -10.43  & -9.66 & -7.67 & -7.18\\ 
\hline  4$\rho_{0}$ & -17.93 & -16.19  & -11.85  & -10.87  \\ 
\hline 
\end{tabular} 
\caption{The mass shifts of $J/\psi$ and $\eta_c$ are shown at densities
of $\rho_0$, 2$\rho_0$ and 4$\rho_0$ at values of the isospin asymmetric
parameter, $\eta$=0 and 0.5 for $\xi$=1. }
\label{table3}
\end{table}

\begin{figure}
\includegraphics[width=18cm,height=18cm]{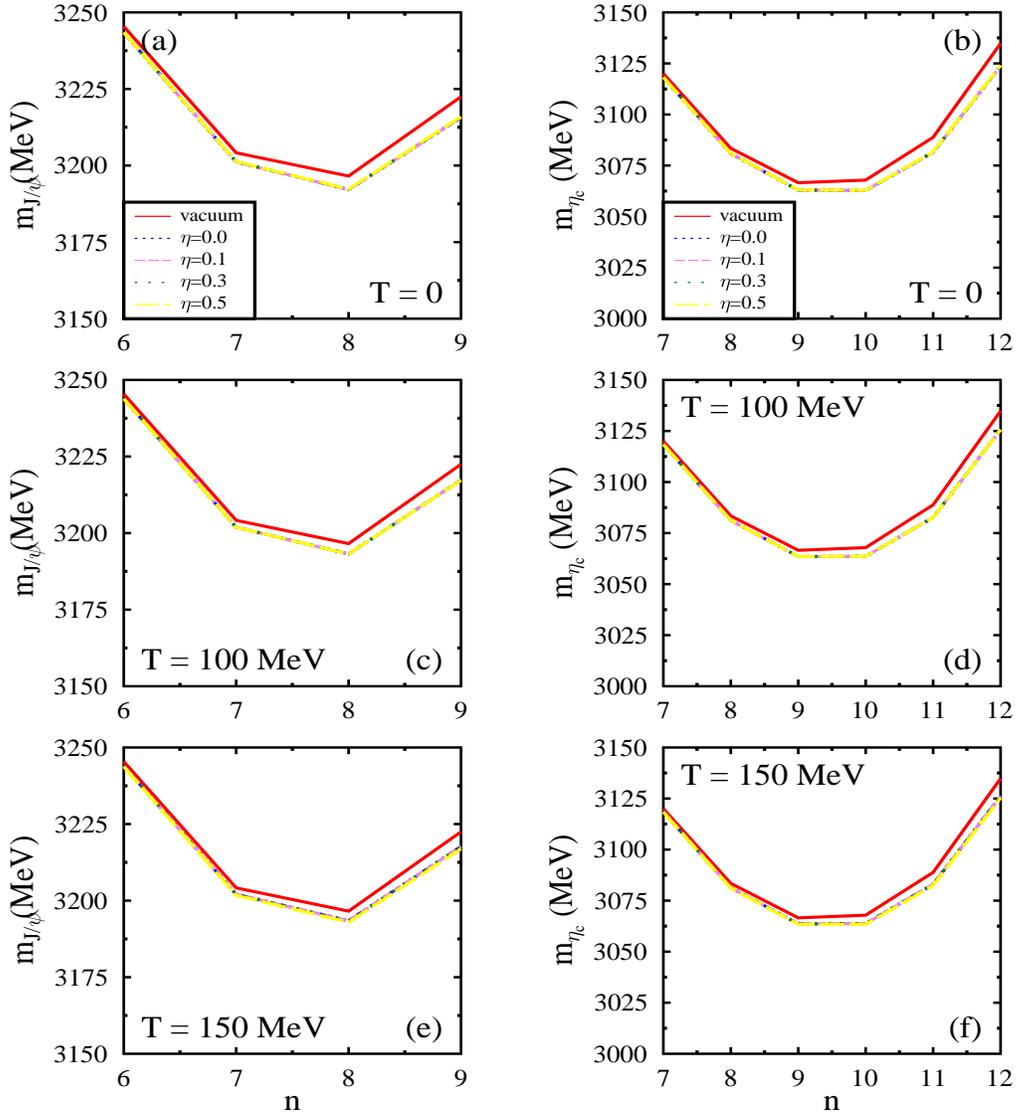} 
\caption{(Color online) The in-medium masses of the $J/\psi$ 
and $\eta_{c}$ mesons plotted as functions of $n$, for  
nuclear matter saturation density, $\rho_{0}$ at different
temperatures and for different values of the isospin asymmetry 
parameter, $\eta$, with $\xi$=1.} 
\label{fig.6}
\end{figure}

\begin{figure}
\includegraphics[width=18cm,height=18cm]{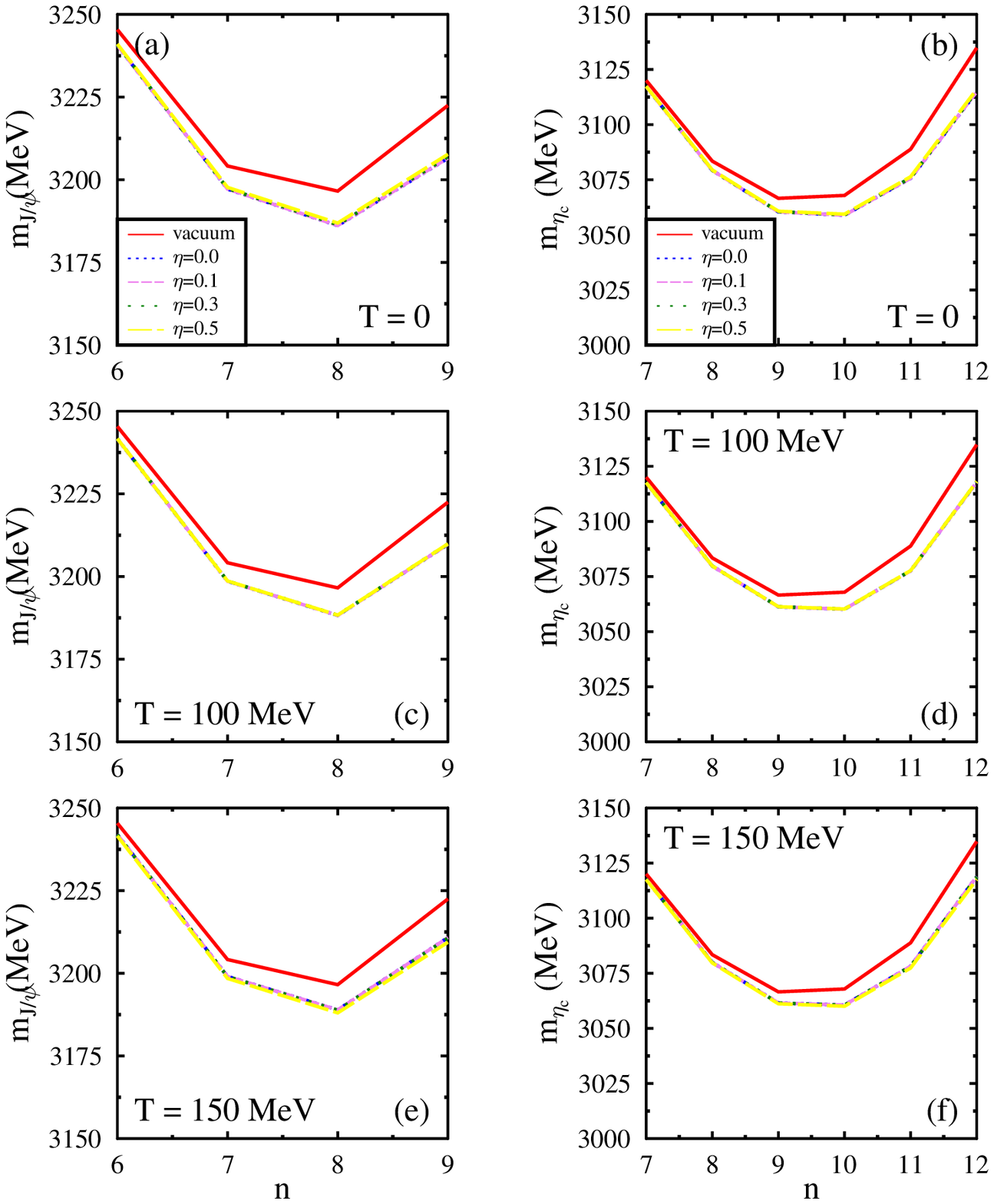} 
\caption{(Color online) The in-medium masses of the $J/\psi$ 
and $\eta_{c}$ mesons plotted as functions of $n$, for  
baryon density of 2$\rho_{0}$, at different
temperatures and for different values of the isospin asymmetry 
parameter, $\eta$, with $\xi$=1.}
\label{fig.7}
\end{figure}

\begin{figure}
\includegraphics[width=18cm,height=18cm]{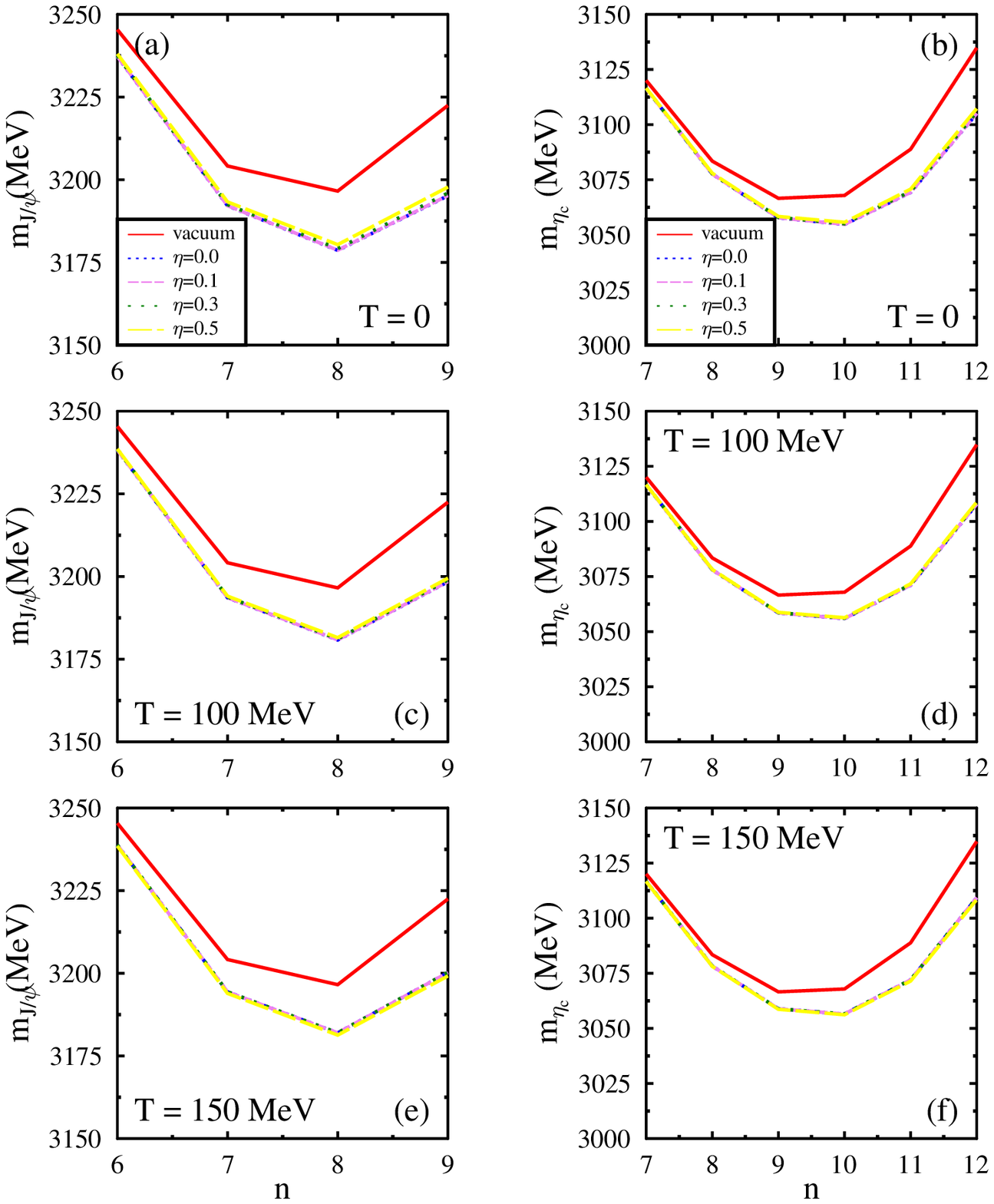} 
\caption{(Color online) The in-medium masses of the $J/\psi$ 
and $\eta_{c}$ mesons plotted as functions of $n$, for  
baryon density of 4$\rho_{0}$ at different
temperatures and for different values of the isospin asymmetry 
parameter, $\eta$, with $\xi$=1.} 
\label{fig.8}
\end{figure}

We also show the results for the mass modifications of $J/\psi$ and 
$\eta_{c}$ mesons, if we consider the value of parameters, $\xi = 1$ 
\cite{klingl}, leading to the values of $\alpha_s$ and $m_c$ as
0.21 and $1.24 \times 10^3$ MeV respectively. Figures \ref{fig.6}, 
\ref{fig.7} and \ref{fig.8} show the temparature and isospin asymmetry
dependence of the mass modifications of $J/\psi$ 
mesons and $\eta_{c}$ mesons, for baryon densities of $\rho_0$,
2$\rho_0$ and 4$\rho_0$ respectively, with parameter $\xi = 1$. 
We observe that with $\xi$=1, the vacuum values of the masses of
$J/\psi$ and $\eta_{c}$ mesons are given as $3196.56$ and 
$3066.57$ MeV respectively. With $\xi$=1, the results for the
mass shifts for $J/\psi$ and $\eta_c$ at different densities
with $\eta$=0 and 0.5, and zero temperature, obtained in the 
present investigation are summarized in table \ref{table3}.
The values of the mass shifts for $J/\psi$ meson in isospin symmetric medium, 
with $\xi = 1$, at nuclear saturation denstiy $\rho_{B} = 
\rho_{0}$ are observed to be $-3.92$, $-3.38$ and $-3.1$ MeV 
for $T = 50, 100$ and $150$ MeV respectively. At baryon density 
$\rho_{B} = 4\rho_{0}$, 
these values of the mass shift change to $-17.23, -15.77$ and 
$-14.59$ MeV at temperature $T = 50, 100$ and $150$ MeV respectively. 
For pseudoscalar meson $\eta_{c}$, the mass shifts at $\rho_{B} = \rho_{0}$ 
are obtained to be $-3.42$, $-3.06$ and $-2.91$ MeV for $T = 50, 100$ 
and $150$ MeV respectively, whereas at $\rho_{B} = 4\rho_{0}$ 
these values of mass shift are seen to be modified to $-11.47$, 
$-10.68$ and $-10.04$ MeV respectively.


In Ref.\cite{kimlee} the operator product expansion was carried out upto 
dimension six and mass shift for $J/\psi$ was found to be $-4$ MeV at 
nuclear saturation density $\rho_{0}$ and temperature $T = 0$. The effect 
of temperature on the $J/\psi$ in deconfinement phase was studied in 
Ref. \cite{leetemp, cesa}. In these investigations, it was reported that 
$J/\psi$ mass is essentially constant in a wide range of temperatures 
and above a particular value of the temperature, T, there is a sharp change 
in the mass of $J/\psi$ in the deconfined phase e.g. in Ref. \cite{lee3} 
the mass shift for $J/\psi$ was reported to be about 200 MeV at 
T = 1.05 T$_{c}$. The pseudoscalar charmonium spectral function 
for different temperatures was studied using a screened potential 
in \cite{mocsy}. The effect of rising temperature was observed to melt
the higher excited states by $1.1T_{C}$ and to shift the continuum
threshold to lower energies. In these studies, it was observed 
that the charmonium $\eta_{c}$ survives even in the deconfined phase.
In Ref.\cite{mocsy,asakawa}, the effect of temperature 
on $\eta_{c}$ in the deconfinement phase was studied. In Ref. 
\cite{asakawa}, it was reported that the $J/\psi$ and $\eta_{c}$ 
survive as distinct resonances in the plasma even upto 
$T \simeq 1.6 T_{c}$ and that they eventually dissociate
between $1.6 T_{c}$ and $1.9 T_{c}$. This suggests that the deconfined 
plasma is non-perturbative enough to hold heavy-quark bound states. 
In the present work, we have studied the effects of temperature on the mass 
modifications of $J/\psi$ and $\eta_{c}$ mesons in the confined phase 
due to modifications of the scalar gluon condensate and twist-2
tnesorial gluon operator, simulated by
a medium dependent scalar dilaton field in chiral SU(3) model and 
the temperature effects are found to be very small as compared to 
the density effects.    

\section{Summary}
\label{sec:5}
In summary, in the present investigation, we have studied the 
mass modifications of the charmonium states, $J/\psi$ and $\eta_{c}$ 
in the nuclear medium using QCD sum rule approach and using modification 
of a dilaton field (which simulates the gluon condensates) within a chiral 
$SU(3)$ model. The in-medium modifications of the $J/\psi$ and $\eta_{c}$ 
are studied as arising due to changes in the scalar and twist-2 gluon 
condensates in the nuclear medium, obtained from the medium modificaion 
of the $\chi$ field. The value of the dilaton field in the hot nuclear
matter is obtained by solving the coupled equations (20) to (23),
which are the equations of motion of $\sigma$, $\zeta$, $\delta$ and 
$\chi$ fields. The dilaton field, $\chi$, thus depends on the scalar 
isovector field, $\delta$, which is related to the isospin asymmetry 
of the nuclear medium. The isospin asymmetry dependence of 
the $\chi$, in turn, leads to the isospin  asymmetry dependence of 
the charmonium states, $J/\psi$ and $\eta_{c}$. The modification of 
the $\chi$ field is observed to be small with the isospin asymmetry 
of the medium, as can be seen from figure \ref{chitemp}. This is 
related to the fact that the magnitude of the obtained value of 
$\delta$ after solving the coupled equations for the scalar fields 
turns out to be much smaller (about few percent) as compared to 
the magnitudes of $\sigma$ and $\zeta$ and hence the isospin 
asymmetry (through $\delta$) only gives rise to a very small modification 
of the dilaton field $\chi$ \cite{amarvind}.
%
It is observed that the temperature effect on the $\chi$ field 
is also very small and the modification of the dilaton field
with density is seen to be the dominant medium effect in the
present investigation. The negligible dependence of the dilaton field 
on isospin asymmetry as well as on temperature is
reflected in the small isospin/temperature dependence of 
the masses of the $J/\psi$ and $\eta_c$ states in the nuclear medium.
Experimentally, 
measurements of dileptons (diphotons) in heavy-ion collisions may provide 
a clue to the properties of vector (pseudo-scalar) mesons in hot/dense 
matter \cite{asakawa}. The present study of the in-medium properties 
of $J/\psi$ and $\eta_{c}$ mesons will be helpful for the experiments 
in the future facility of the FAIR, GSI, where the compressed baryonic 
matter at high densities and moderate temperature will be produced.

\acknowledgements
Financial support from Department of Science and Technology, Government 
of India (project no. SR/S2/HEP-21/2006) is gratefully acknowledged. 
One of the authors (AM) is grateful to FIAS, University of Frankfurt,
for warm hospitality and acknowledges financial support from Alexander
 von Humboldt Stiftung when this work was initiated. 
 

\end{document}